\documentclass[aps,preprint,prl, superscriptaddress, showpacs,nobibnotes]{revtex4-2}
\usepackage{graphicx}
\usepackage{bm,amsmath}
\usepackage{dcolumn}
\usepackage{natbib,xcolor,float}
\usepackage{mhchem,bm}
\usepackage{placeins}
\usepackage{ragged2e}
\makeatletter
\def\@fnsymbol#1{\ensuremath{\ifcase#1\or \dagger\or \dagger,*\or
   \mathsection\or \mathparagraph\or \|\or **\or \dagger\dagger
   \or \ddagger\ddagger \else\@ctrerr\fi}}
   \makeatother

\begin{document}

\title{Ferromagnetic helical nodal line and Kane-Mele spin-orbit coupling in kagome metal \ce{Fe3Sn2}}
\author{Shiang Fang}
\thanks{These authors contributed equally.}
\affiliation{Department of Physics and Astronomy, Center for Materials Theory, Rutgers University, Piscataway, New Jersey 08854, USA}
\author{Linda Ye}
\thanks{These authors contributed equally.}
\homepage{Present Address: Department of Applied Physics, Stanford University, Stanford, California 94305, USA}
\affiliation{Department of Physics, Massachusetts Institute of Technology, Cambridge, Massachusetts 02139, USA}
\author{Madhav Prasad Ghimire}
\affiliation{Central Department of Physics, Tribhuvan University, Kirtipur, 44613, Kathmandu, Nepal}
\affiliation{Leibniz Institute for Solid State and Materials Research, IFW Dresden, Helmholtzstraße 20,  01069 Dresden, Germany}
\author{Min Gu Kang}
\affiliation{Department of Physics, Massachusetts Institute of Technology, Cambridge, Massachusetts 02139, USA}
\author{Junwei Liu}
\affiliation{Department of Physics, Hong Kong University of Science and Technology, Clear Water Bay, Hong Kong, China}
\author{Liang Fu}
\affiliation{Department of Physics, Massachusetts Institute of Technology, Cambridge, Massachusetts 02139, USA}
\author{Manuel Richter}
\affiliation{Leibniz Institute for Solid State and Materials Research, IFW Dresden, Helmholtzstraße 20, 01069 Dresden, Germany}
\affiliation{Dresden Center for Computational Materials Science (DCMS), TU Dresden,  01062 Dresden, Germany}
\author{Jeroen van den Brink}
\affiliation{Leibniz Institute for Solid State and Materials Research, IFW Dresden, Helmholtzstraße 20, 01069 Dresden, Germany}
\affiliation{W\"urzburg-Dresden Cluster of Excellence ct.qmat, Technische Universit\"at Dresden, 01062 Dresden, Germany}
\author{Efthimios Kaxiras}
\affiliation{Department of Physics, Harvard University, Cambridge, Massachusetts 02138, USA}
\affiliation{John A. Paulson School of Engineering and Applied Sciences, Harvard University, Cambridge, Massachusetts 02138, USA}
\author{Riccardo Comin}
\affiliation{Department of Physics, Massachusetts Institute of Technology, Cambridge, Massachusetts 02139, USA}
\author{Joseph G. Checkelsky}
\thanks{checkelsky@mit.edu}
\affiliation{Department of Physics, Massachusetts Institute of Technology, Cambridge, Massachusetts 02139, USA}
\date{\today}

\begin{abstract}
\textbf{
The two-dimensional kagome lattice hosts Dirac fermions at its Brillouin zone corners K and K', analogous to the honeycomb lattice. In the density functional theory electronic structure of ferromagnetic kagome metal \ce{Fe3Sn2}, without spin-orbit coupling we identify two energetically split helical nodal lines winding along $z$ in the vicinity of K and K' resulting from the trigonal stacking of the kagome layers. We find that hopping across A-A stacking introduces a layer splitting in energy while that across A-B stacking controls the momentum space amplitude of the helical nodal lines. The effect of spin-orbit coupling is found to resemble that of a Kane-Mele term, where the nodal lines can either be fully gapped to quasi-two-dimensional massive Dirac fermions, or remain gapless at discrete Weyl points depending on the ferromagnetic moment orientation. Aside from numerically establishing \ce{Fe3Sn2} as a model Dirac kagome metal, our results provide insights into materials design of topological phases from the lattice point of view, where paradigmatic low dimensional lattice models often find realizations in crystalline materials with three-dimensional stacking.
}
\end{abstract}

\maketitle
\subsection{Introduction}
Topological nodal lines are one-dimensional manifolds of band degeneracies in momentum space first introduced conceptually by Burkov \textit{et al.} in 2011 as a higher dimensional generalization of point-like band touching \cite{Burkov}. Such line nodes have in recent years found realizations in various forms in crystalline materials, including infinite lines extending over Brillouin zones \cite{AlB2}, closed loops \cite{Ca3P2}, along with intricate three dimensional networks of chains, knots, and nexuses \cite{Chain,Co2MnGa,nexus,nonAbelian,Fang_review,Yan_review}. Interest in electronic line nodes are partly motivated by the peculiar emergent condensed matter quasiparticles they support, which do not possess fundamental particle analogues \cite{Chain}. Furthermore, due to bulk-boundary correspondence, nodal lines in three dimensional bulk materials generate surface states enclosed in their surface projection; these signature surface states under certain circumstances bear little  momentum-space dispersion over a finite region in the surface Brillouin zone \cite{Burkov} and are therefore termed ``drumhead surface states". The enhanced density of states of drumhead surface states is expected to provide a route towards high-temperature correlated phases including ferromagnetism and superconductivity \cite{Graphite_JETP}.

Viewed in the context of band topology, nodal lines in three-dimensional materials are necessarily protected by symmetries \cite{Fang_review} and therefore serve as progenitors for a large number of distinct topological electronic states when the corresponding symmetry is relieved. For instance, broken mirror symmetry is suggested to separate intersecting nodal lines and serve to manipulate an embedded non-Abelian topology \cite{nonAbelian}. The prototype inversion symmetry- breaking Weyl semimetal TaAs \cite{TaAs} and time-reversal symmetry breaking Weyl semimetal \ce{Co3Sn2S2} \cite{Co3Sn2S2} can both be viewed as generated by adding spin-orbit coupling--which breaks the SU(2) spin-rotation symmetry--to nodal loops on mirror planes.  In addition, nodal lines in certain cases can be fully gapped and further give rise to topological insulating phases \cite{Yan_review}. From the materials perspective, elucidating mechanisms of generating topological nodal lines and their interplay with different types of symmetries--including crystallographic symmetries, spin-rotation symmetry, and time-reversal symmetry--are expected to afford key clues in discovering novel topological electronic states and allow the study of emergent electromagnetic responses in such systems.
 
Motivated by the experimental discovery of quasi-two-dimensional Dirac electronic dispersions in the vicinity of the Fermi level in the ferromagnetic kagome metal \ce{Fe3Sn2} \cite{Fe3Sn2_ARPES}, we here examine the density functional theory (DFT) electronic structure of the system in the context of three-dimensional (3D) topological nodal lines. In the following, we use the convention of Dirac fermions referring to linearly dispersing two-dimensional bands as in topological insulator surface states \cite{TI} or the two-dimensional graphene \cite{Graphene} and kagome models \cite{kagome_Hoffman}. For the case of the considered 3D material, this convention includes crossing states with linear dispersion in two dimensions and preserved degeneracy in the third dimension (nodal lines), regardless of their degeneracy. This should be contrasted with the convention associated with the four-fold degeneracy of 3D Dirac semimetals that are described by the 3D Dirac equation \cite{DSM}. The two-dimensional kagome lattice is composed of corner-shared triangles (see Fig. \ref{fig-1}(a)) and is known theoretically to host Dirac fermions at its Brillouin zone corners K and K' in the electronic spectrum as illustrated in Fig. \ref{fig-1}(b) -- analogous to the honeycomb lattice \cite{kagome_Hoffman}. In contrast to the honeycomb lattice whose experimental realization primarily falls into $p$-electron materials such as graphene and other main group X-enes \cite{graphene_analogue}, the kagome lattice has found extensive presence in a class of transition metal intermetallic compounds termed ``kagome metals", where the kagome bands are composed by $d$ electrons \cite{Fe3Sn2_ARPES,FeSn,Co3Sn2S2,CoSn,Mn3Ge,Mn3Sn,KV3Sb5,TbMn6Sn6}. As these compounds crystallize in three-dimensional structures, a natural question is how the notion of the point nodes in the two-dimensional limit can be extended to the third dimension. The subject of this study -- binary ferromagnetic kagome metal \ce{Fe3Sn2} has been experimentally identified as host of bulk quasi-two-dimensional Dirac fermions in transport and photoemission spectroscopy \cite{Fe3Sn2_ARPES}, as well as in de Haas-van Alphen quantum oscillations \cite{Fe3Sn2_dHvA} and optical conductivity \cite{Fe3Sn2_optical}. Scanning tunneling microscopy has revealed a strongly anisotropic response of the electronic structure of \ce{Fe3Sn2} due to spin-orbit coupling \cite{Fe3Sn2_STM}, and more recently a large number of Weyl points are also proposed to be present in the system \cite{Fe3Sn2_Weyl}. In view of the successful application of DFT to related topological kagome metals \cite{FeSn,Mn3Ge}, a  comprehensive DFT study of the electronic structure of \ce{Fe3Sn2} is expected to address the nature of its electronic topology and offer insights into the origin of experimentally observed Dirac fermions. 

In this study we first identify two sets of ferromagnetic helical nodal lines in \ce{Fe3Sn2} near K and K' of the hexagonal Brillouin zone in the limit of vanishing spin-orbit coupling. We also found that with the introduction of spin-orbit coupling, these nodal lines are gapped into a three-dimensional quantum anomalous Hall insulating phase with out-of-plane ferromagnetic moments while with in-plane moments, point Weyl nodes remain gapless along the helices. The helical nodal lines are found to originate from the rhombohedral stacking of the bilayer kagome lattices and also subject to a layer splitting between upper and lower branches. We propose that these ferromagnetic helical nodal lines are the key to describe the topological electronic structure in \ce{Fe3Sn2} and insights obtained herein can be broadly applied to various three dimensional constructions of two-dimensional lattice models. 

\subsection{Ferromagnetic helical nodal line in \ce{Fe3Sn2}}

\begin{figure}[h]
	\includegraphics[width = 0.8\columnwidth]{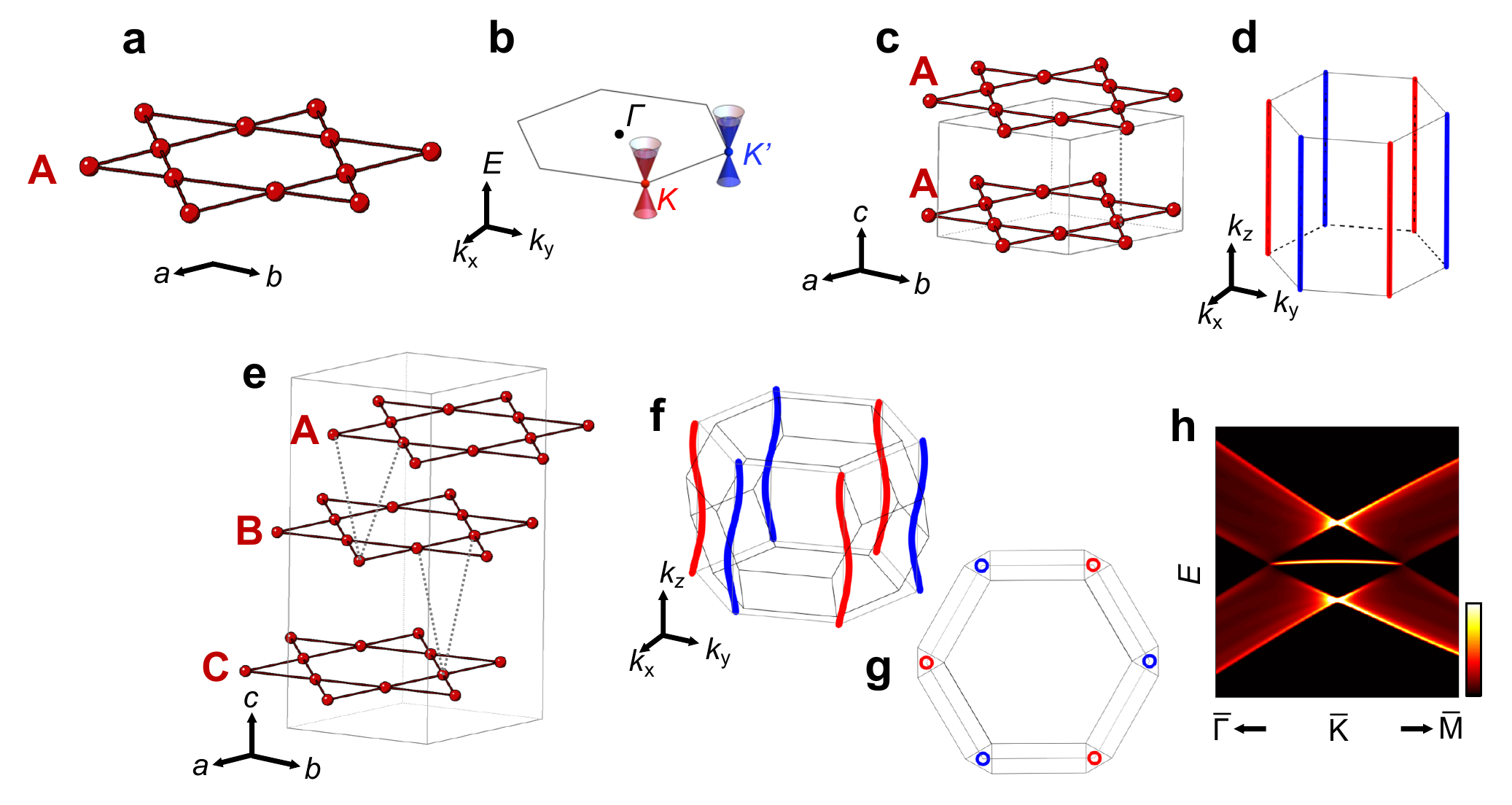}
	\caption{\label{fig-1} \textbf{Nodal lines and three dimensional stacking of the kagome lattice}
	(a) The two dimensional (2D) kagome lattice and (b) associated Dirac fermions in the hexagonal Brillouin zone (BZ). Blue and red Dirac fermions at $\mathrm{K}$ and $\mathrm{K}'$ possess opposite chiralities. (c) Schematic of a three dimensional (3D), A-A stacked kagome lattice and (d) the corresponding vertical nodal lines in the hexagonal prism BZ. (e) Schematic of an A-B-C stacking of the kagome lattice and the corresponding helical nodal lines are shown in (f) and (g) from both an isometric (f) and a top (g) perspective. In (c) and (e) the dashed lines represent the interplane hopping $t_1$ while the in-plane kagome bonds are characterized by an hopping integral $t_0$. In (f) we show the rhombohedral BZ along with a hexagonal prism extended from the 2D BZ illustrated in (b). (h) Surface spectra weight of the A-B-C kagome tight-binding model. A drumhead surface state (DSS) can be identified as the flat and bright intensity within the projection of the helical nodal line to the surface BZ.}
\end{figure}

 We start from the two-dimensional (2D) nearest neighbor tight-binding model of the kagome lattice as shown in Fig. \ref{fig-1}(a), where we highlight the Dirac fermions located at $\mathrm{K}$ and $\mathrm{K}'$, with opposite chirality shown in red and blue, respectively in Fig. \ref{fig-1}(b). In Fig. \ref{fig-1}(c-g) we illustrate the effects of three-dimensional (3D) stacking with a moderate interplane hopping on these Dirac fermions. For clarity in Fig. \ref{fig-1}(c,e) we introduce only the in-plane nearest neighbor hopping $t_0$ (solid lines) and the nearest out-of-plane hopping $t_1$ (dashed lines). The location of band crossing points in the 3D Brillouin zone (BZ) with $t_0=1,t_1=0.1$ are shown in Fig. \ref{fig-1}(d,f,g). In the simpler A-A stacking the hopping along the $z$ direction extends the Dirac points at $\mathrm{K}$ and $\mathrm{K}'$ to vertical nodal lines along the $\mathrm{K-H}$ ($\mathrm{K'-H'}$) directions (Fig. \ref{fig-1}(d)). A-B-C stacking instead results in helical nodal lines where at each $k_z$ plane the Dirac points are shifted away from $\mathrm{K}$ and $\mathrm{K'}$ (Fig. \ref{fig-1}(f,g)). Here we show the BZ of the rhombohedral unit cell of the A-B-C stacking in Fig. \ref{fig-1}(f,g) along with a hexagonal prism extended vertically from the original 2D hexagonal BZ; the projected helical nodal lines wind around the corresponding $\mathrm{K}$ and $\mathrm{K}'$ points of the latter. We show in Fig. \ref{fig-1}(g) a top view of the helices, where within the projection onto the top surface, a weakly dispersive drumhead surface state (DSS) can be found in the surface spectral function using a large finite slab calculation (Fig. \ref{fig-1}(h)), as is expected for prototypical topological nodal line semimetals \cite{Fang_review}. We note that similar helical nodal lines have been discussed in the context of A-B-C stacked rhombohedral graphite \cite{ABC_Graphite_1969,Graphite_IFW,ABC_Graphite_2016,Graphite_JETP}; there the associated drumhead surface states are theoretically anticipated to drive correlated magnetic and superconducting states \cite{Graphite_JETP,Graphite_otani} and have been observed in photoemission spectroscopy in multi-layer A-B-C stacked graphite flakes \cite{Graphite_ARPES}. In the context of the kagome lattice, it is intriguing to note that the three-dimensional stacking allows access to a surface flat band and potential correlated states it entails, in addition to the in-plane destructive interference-induced flat band of bulk nature \cite{CoSn,BalentsFB}. 

\begin{figure}[p]
	\includegraphics[width = 0.8\columnwidth]{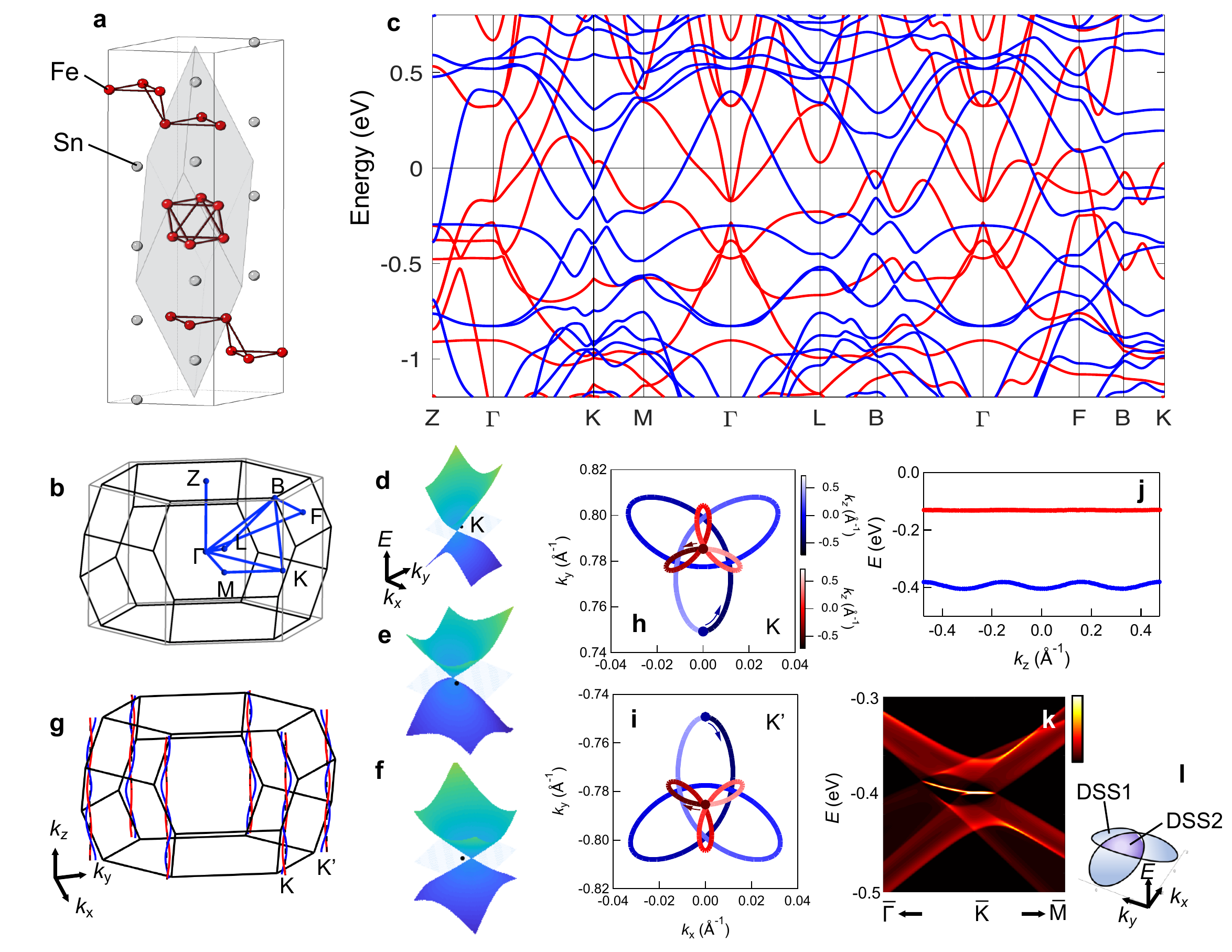}
	\caption{\label{fig-2} \textbf{Scalar-relativistic electronic structure and helical nodal lines in \ce{Fe3Sn2}}
	(a) Crystal structure of \ce{Fe3Sn2} with iron atoms shown in red and tin atoms in gray. The primitive rhombohedral unit cell containing a kagome bilayer is outlined in gray. (b) Schematic of the rhombohedral BZ of \ce{Fe3Sn2} with high symmetry points labeled and high symmetry directions highlighted in blue. The gray hexagonal prism is extended from the hexagonal BZ in the 2D limit. (c) Scalar-relativistic generalized gradient approximation (GGA) DFT electronic structure of \ce{Fe3Sn2} where the majority spin is shown in red and minority in blue. (d-f) DFT energy-momentum dispersion of the upper helical nodal line within a region of $0.16\times0.16\mathrm{\AA}^{-2}$ close to $\mathrm{K}$ at (d) $k_z=0$, (e) $k_z=\dfrac{2\pi}{3c}$, (f) $k_z=\dfrac{4\pi}{3c}$ planes, respectively. The black sphere denotes the $\mathrm{K}$ point of the 2D BZ. (g) The helical nodal lines around $\mathrm{K}$ and $\mathrm{K}$' in \ce{Fe3Sn2}; the upper nodal line is shown in red and lower nodal line in blue. (h,i) Magnified top view of the helical nodal lines at $\mathrm{K}$ (h) and $\mathrm{K}$' (i), respectively. The color gradient in (h,i) reflects the value of $k_z$. (j) Energy dispersion of both the upper and lower nodal lines along $k_z$. (k) Surface spectra weight of the lower nodal line inferred from the $\bm{k}\cdot\bm{p}$ model (see Methods). (l) Schematic of two distinct momentum regions DSS1 (blue) and DSS2 (purple) that host different number of drumhead surface states. The hypotrochoid curve represents the projection of lower nodal line to the top surface.}
\end{figure}

Having illustrated the generation of helical nodal lines in a simple A-B-C stacked kagome lattice model, in the following we turn to the DFT electronic structure of \ce{Fe3Sn2} in the absence of spin-orbit coupling -- to test the relevance of the above picture in describing the system. The crystalline structure of \ce{Fe3Sn2} (Space group No.166 $R\bar{3}m$) is illustrated in Fig. \ref{fig-2}(a) in the conventional hexagonal unit cell, while the rhombohedral unit cell is highlighted in gray. Each unit cell contains a bilayer kagome structure that are further stacked in the A-B-C fashion. In Fig. \ref{fig-2}(b) we show the rhombohedral BZ of \ce{Fe3Sn2} along with selected high symmetry points. We note that in addition to $\mathrm{Z}$, $\mathrm{\Gamma}$, $\mathrm{B}$, $\mathrm{L}$, $\mathrm{F}$ of the rhombohedral convention, we also include $\mathrm{M}$ and $\mathrm{K}$ of the hexagonal convention to better describe the Dirac electronic structure observed experimentally in the proximity of $\bar{\mathrm{K}}$ of the surface BZ \cite{Fe3Sn2_ARPES}. The calculated electronic structure is shown in Fig. \ref{fig-2}(c) along the high symmetry lines highlighted in Fig. \ref{fig-2}(b). The majority spin states (illustrated in red) feature electron pockets centered near $\mathrm{\Gamma}$ and a hole pocket close to $\mathrm{K}$, while the minority spin states (illustrated in blue) show a double Dirac structure displaced in energy in the vicinity of $\mathrm{K}$, as previously observed in angle-resolved photoemission experiments \cite{Fe3Sn2_ARPES,Fe3Sn2_Weyl,Fe3Sn2_Kondo}. This suggests that DFT reasonably account for the electronic structure in \ce{Fe3Sn2}. Hereafter we refer to the Dirac structure near -0.1 eV (-0.4 eV) as upper (lower) Dirac fermions, respectively.

Focusing on the upper Dirac dispersion, although we observe an apparent gap at $\mathrm{K}$ similar to the DFT band structure reported in Ref. \cite{Fe3Sn2_Kondo}, via searching in the proximity of $\mathrm{K}$ we find the gap closing and reopening through a single point at each constant $k_z$ cross-section (Band landscapes at selected $k_z$ planes for the upper nodal line are shown in Fig. \ref{fig-2}(d-f)). A search near the lower Dirac dispersion yields similar results. Connecting the point nodes at each $k_z$ plane we obtain two sets of helical nodal lines as depicted in Fig. \ref{fig-2}(g), where the nodal line for the upper Dirac fermion is shown in red, and lower Dirac fermions in blue. Both nodal lines wind around $\mathrm{K}$ vertical in a helical fashion, similar to that of the simple tight-binding model as shown in Fig. \ref{fig-1}(f,g), suggesting that the trigonal A-B-C stacking plays a key role in generating the helical nodal lines. A magnified view of the top projection of the helical nodal lines at $\mathrm{K}$ and $\mathrm{K}$' can be found in Fig. \ref{fig-2}(h) and (i) in the shape of hypotrochoids, where we use gradient scales as shown in Fig. \ref{fig-2}(h) to sketch the evolution along $k_z$. Near $\mathrm{K}$ both nodal lines may be approximately described by the following functional form:
\begin{equation}
    \Delta k_x + i\Delta k_y= i\lambda_1 e^{-i k_z c}+i\lambda_2 e^{2 i k_z c}
\end{equation} Here the band touching point is shifted to $(\Delta k_x,\Delta k_y)$ with respect to K point at given $k_z$. For the upper (lower) nodal line, the parameters are $\lambda_1^u=0.00928$ ($\lambda_1^l=0.0136$) and $\lambda_2^u=0.0087$ ($\lambda_2^l=-0.0214$) in units of \AA$^{-1}$, respectively. Here $c$ stands for the vertical distance between the kagome bilayer units; superscripts $u$ and $l$ stand for upper and lower nodal lines, respectively. The $k_z$-evolution near $\mathrm{K}$' can be obtained by performing an inversion operation to that near $\mathrm{K}$ (Fig. \ref{fig-2}(i)).

The presence of sinusoidal components of both $k_zc$ and $2k_zc$ implies the presence of both nearest layer and next nearest layer hopping terms in \ce{Fe3Sn2} (see Supplementary Materials), the latter not included in the simple nearest layer model discussed above in Fig. \ref{fig-1}(f-h). We further examine the energy evolution of both the upper and lower nodal lines in Fig. \ref{fig-2}(j). The closer confinement of the upper nodal lines to the verticals of $\mathrm{K}$ and $\mathrm{K}$' is accompanied by a weaker out-of-plane dispersion illustrated in Fig. \ref{fig-2}(j): the energy variation of the upper nodal line is on the order of 1.5 meV while for the lower is on the order of 11 meV, while both are significantly weaker as compared to the in-plane Dirac bandwidth $\sim2$ eV. This corroborates the bulk quasi-2D nature and the absence of photon-energy dependence of the double Dirac structure observed in \ce{Fe3Sn2} \cite{Fe3Sn2_ARPES}.

To further elucidate the nature of the identified helical nodal lines, we have computed the Berry phase $\Phi_B=\oint_{\Gamma_B}\bm{A}_{\bm{k}}\cdot d\bm{k}$ on loops $\Gamma_B$ around the nodal lines where $\bm{A}_{\bm{k}}$ is the Berry connection $\bm{A}_{\bm{k}} = -i \langle u_{\bm{k}}|\nabla_{\bm{k}}|u_{\bm{k}}\rangle$ \cite{Mikitik}, with $|u_{\bm{k}}\rangle$ denoting the wave function at $\bm{k}$. Without spin-orbit coupling, a combined inversion and effective time-reversal symmetry dictates the quantization of the Berry phase as a binary $Z_2$ invariant that takes either 0 or $\pi$ \cite{Fang_review}. We have verified that both upper and lower nodal lines support a $\pi$-Berry phase to the path integrals enclosing the nodal lines, suggesting that it is the the non-trivial Berry phase that protects the nodal lines in the present case. The $\pi$-Berry phase here may be naturally connected to that of Dirac fermions in the 2D limit \cite{Berry_graphene} and that more recently demonstrated in photoemission intensity analysis for bulk quasi-2D Dirac fermions derived from the kagome lattice in FeSn \cite{FeSn}. In the present system, due to a reduction of the symmetry from hexagonal to trigonal, the positions of the nodal lines are displaced from high symmetry lines; nevertheless their presence is robust and protected by the $\pi$-Berry phase inherited from the 2D limit. In Fig. \ref{fig-si-nl-hopping} we show that nodal lines centered at $\mathrm{K}$ and $\mathrm{K'}$ are robust with increasing interplane hopping strength, as long as the two lines do not touch and hybridize with each other. We note that due to broken time reversal symmetry, each band touching point here in \ce{Fe3Sn2} is ferromagnetic and two-fold degenerate, which belongs to a similar class with the ferromagnetic nodal lines discussed in \ce{Co2MnGa} \cite{Co2MnGa}, \ce{Co3Sn2S2} \cite{Co3Sn2S2} and \ce{Fe3GeTe2} \cite{Fe3GeTe2} in the absence of spin-orbit coupling.

Additionally, we examine the surface states which originate from the hypotrochoid winding pattern of the lower nodal line in Fig. \ref{fig-2}(k,l) (see Methods). The existence of a drumhead surface state in a nodal line semimetal may be illustrated in the following picture: for a given $(k_x,k_y)$ one may define a Zak phase $\Phi_Z=\int_{-\pi/c}^{\pi/c}\mathbf{A}\cdot dk_z$ along $k_z$, and $\Phi_Z(k_x,k_y)=\pi$ corresponds to a 1D topological insulator with zero energy edge states (here we restrict ourselves in the spinless case) and defines the  $(k_x,k_y)$ region where surface states reside \cite{Ca3P2}. In Fig. \ref{fig-2}(k), we find that in the present case ferromagnetic drumhead surface states appear once within the three side lobes (blue region labeled DSS1 in Fig. \ref{fig-2}(l)) with $\Phi_Z(k_x,k_y)=\pi$ and twice within the center surface momentum regime (purple region labeled DSS2 in Fig. \ref{fig-2}(l)) $\Phi_Z(k_x,k_y)=2\pi (0)$. We expect the surface states in the DSS2 region to be more fragile and dependent on the surface potential than that within DSS1, as has been discussed for systems with multiple nodal loops \cite {Multiple_NL}. The lobe structure of the flat surface bands adds new opportunties for potential correlated phenomena; moreover, with the sensitivity of the helical nodal line to interplane hopping, one may manipulate the connectivity and drive Lifshitz transitions of these lobe-wise drumhead surface states by hydrostatic pressure (see Fig. \ref{fig-si-pressure}).

\subsection{Kane-Mele spin-orbit coupling in \ce{Fe3Sn2}}

\begin{figure}[h]
	\includegraphics[width = \columnwidth]{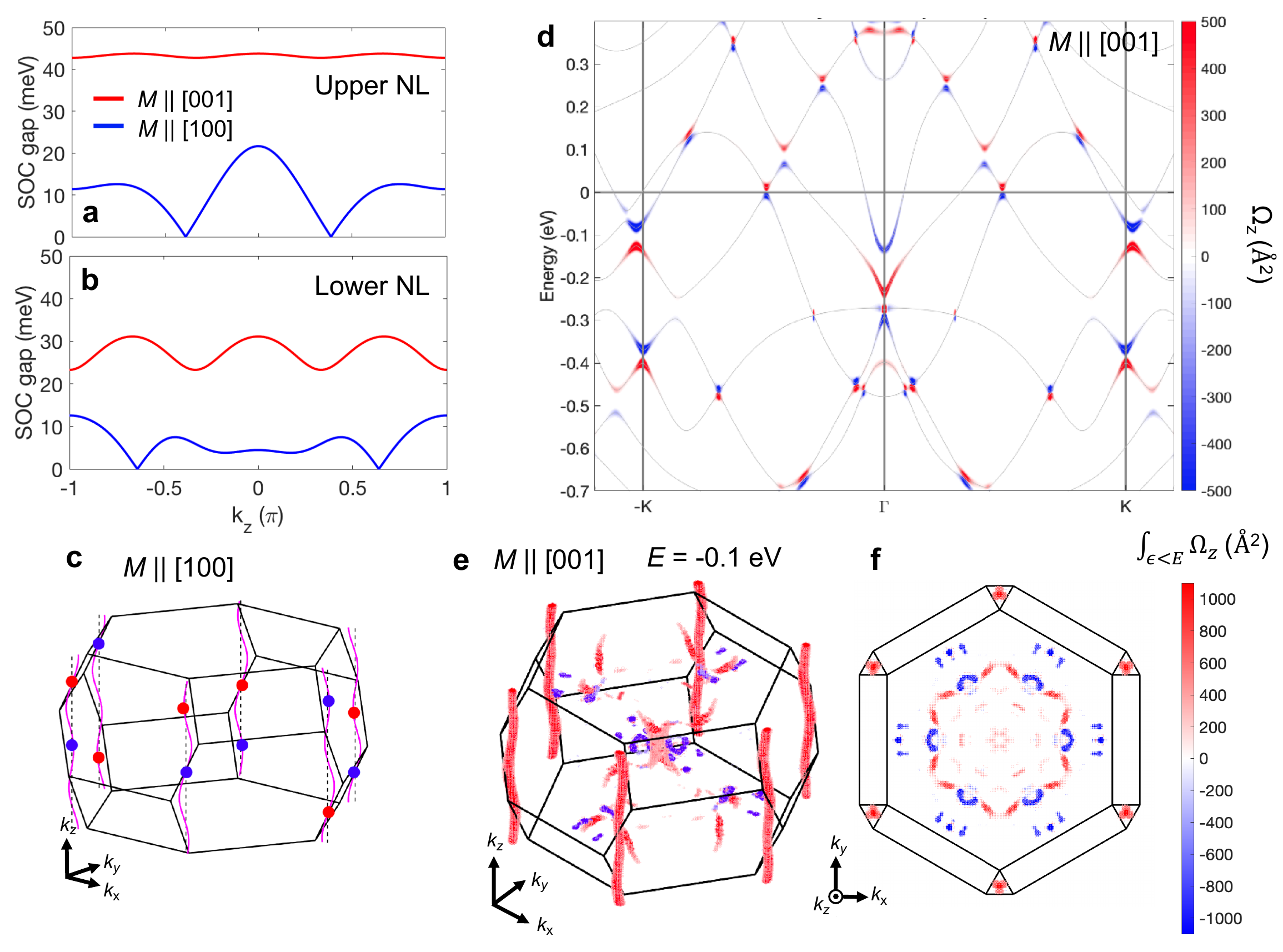}
	\caption{\label{fig-3} \textbf{Electronic structure of \ce{Fe3Sn2} with spin-orbit coupling}
(a) The gap along the helical nodal line at each $k_z$ with ferromagnetic moment out-of-plane (red) and in-plane along [100] (blue) for the upper nodal line. (b) A similar analysis for the lower nodal line. (c) Weyl points originated from the lower helical nodal line with magnetic moment along [100]. The nodal line itself is shown in magenta. (d) The Berry curvature $\Omega_z$ distribution with the ferromagnetic moments pointing out-of-plane along $\mathrm{\Gamma}$ - $\mathrm{K}$ - $\mathrm{K}$' high symmetry line in the band structure. (e,f) Distribution of integrated Berry curvature (see text) up to the upper Dirac gap in the 3D BZ at $E=-0.1$ eV in a 3D view (e) and top view (f), respectively.}
\end{figure}

Having located the helical nodal lines in the proximity of $\mathrm{K}$ and $\mathrm{K}$' in the absence of spin-orbit coupling, in the following we examine the fully relativistic electronic structure of \ce{Fe3Sn2}. As \ce{Fe3Sn2} is known to be a soft ferromagnet \cite{Fe3Sn2_neutron}, we consider both cases of moments in and out of the kagome lattice plane. With an out-of-plane magnetic moment, we find that both the upper and lower nodal lines are fully gapped with spin-orbit coupling, with the upper $(43.3\pm0.5)$ meV and lower Dirac gap $(27\pm4)$ meV as shown in Fig. \ref{fig-3}(a,b). The in-plane magnetic moment induces a smaller gap, and the nodal lines remain gapless at two $k_z$ positions for both the upper and lower branches (Fig. \ref{fig-3}(a,b)). This anisotropic coupling of $\bm{M}$ with the Dirac electrons is consistent with a Kane-Mele type spin-orbit coupling in \ce{Fe3Sn2} as suggested in Refs. \cite{Fe3Sn2_ARPES,Fe3Sn2_dHvA}. These discrete remnant touching points correspond to Weyl points; we show Weyl points with the opposite chirality as blue and red circles in Fig. \ref{fig-3}(c) for the case of the lower nodal line.

Next we elaborate on the out-of-plane ferromagnetic case where spin-orbit coupling introduces a full gap to the helical nodal lines. We have computed the resultant Berry curvature $\bm{\Omega}=\bm{\nabla}\times\bm{A}_{\bm{k}}$ and its distribution along high symmetry lines $\mathrm{K}$'-$\mathrm{\Gamma}$-$\mathrm{K}$ in $k_z=0$ plane is shown in Fig. \ref{fig-3}(d). We observe concentrated Berry curvature $\Omega_z$ at the gapped nodes as expected for massive Dirac fermions \cite{graphene_inversion}, together with additional distribution of Berry curvatures from potential Weyl points (see also Fig. \ref{fig-si-weyl}) \cite{Fe3Sn2_Weyl}. $\Omega_z$ near $\mathrm{K}$ and $\mathrm{K}$' are found to be additive, which can be contrasted to the cancelling $\Omega_z$ pattern at $\mathrm{K}$ and $\mathrm{K}$' valleys in the inversion-symmetry-breaking and time-reversal-symmetric graphene \cite{graphene_inversion}. The Berry curvature structure at both upper and lower Dirac gaps also exhibit the same sign. We further illustrate the distribution of integrated Berry curvature $\int_{\epsilon<E}\Omega_z^{\epsilon}$ (here $\epsilon$ represents all states with energy below $E$) in the 3D BZ in Fig. \ref{fig-3}(e,f) up to the upper Dirac gap ($E=-0.1$ eV), where columns of Berry curvature hot spots are confined along the stacked massive Dirac fermions. These massive Dirac fermions descend naturally from their 2D limit as kagome realizations of the Haldane model \cite{Haldane} and therefore in isolation form a 3D quantum anomalous Hall insulating phase \cite{MTI_layer}. In this context, we propose that chiral boundary modes can be detected at step edges of the kagome cleavage of \ce{Fe3Sn2} crystals at energies within the Dirac gap, similar to those recently demonstrated in a Mn-based kagome metal \ce{TbMn6Sn6} \cite{TbMn6Sn6}. Near $\mathrm{\Gamma}$ we also observe less extended patches of Berry curvature intensities, the 3D nature of which suggests that they may originate from underlying Weyl fermions in the system \cite{Fe3Sn2_Weyl}. The difference in momentum-space dimensionality leads to a response dominated by the extended $k_{z}$ features associated with the massive quasi-2D Dirac states (Fig. \ref{fig-3}(e)).

Having demonstrated that nodal lines gapped by the interplay of ferromagnetic order and spin-orbit coupling in \ce{Fe3Sn2} serve as a strong source of Berry curvature and therefore contribute significantly to the intrinsic anomalous Hall conductivity $\sigma_{xy}$ (see Fig. \ref{fig-si-sxy})\cite{Fe3Sn2_ARPES}, it is instructive to compare the helical nodal lines identified here in \ce{Fe3Sn2} with the nodal lines discussed in the van der Waals ferromagnet \ce{Fe3GeTe2} \cite{Fe3GeTe2}. In both cases, topological nodal lines are rendered strong sources of $\Omega_z$. We note that as compared with \ce{Fe3GeTe2}, where contribution to $\sigma_{xy}$ is concentrated in the momentum space near the gapped nodal line along $\mathrm{K-H}$ over its energy dispersion of 0.25 eV \cite{Fe3GeTe2}, such contributions in \ce{Fe3Sn2} is further concentrated energetically due to the weak energy variation of the massive Dirac fermions along the $z$ direction. Intriguingly, in \ce{Fe3GeTe2}, it is also found that an out-of-plane moment maximizes the spin-orbit gap along the nodal line. Despite this similar sensitivity with the ferromagnetic moment orientation, we note that the spin-orbit coupling in \ce{Fe3Sn2} that opens the gaps at the Dirac nodes is different at the effective model level than that discussed for \ce{Fe3GeTe2} \cite{Fe3GeTe2}. In the latter, an on-site spin-orbit coupling of the $\bm{L}\cdot\bm{S}$ form lifts the degeneracy at $\mathrm{K}$ that originates from orbital degrees of freedom of Fe $d$-orbitals; this mechanism also applies to the $p_x,p_y$ models on the triangular lattices \cite{triangular_pxpy} and the $d$ orbitals on hexagonal closed packing cobalt layers \cite{hcp_Co}, where an orbital degree of freedom is preserved for three-(or six-)fold rotation centers. In the context of the kagome lattice, the degeneracies of all $d$ orbitals are in principle lifted due to the low site symmetry. An onsite spin-orbit coupling term is therefore ineffective in opening a gap for the band crossing at the effective model level, rather the intersite form of spin-orbit coupling -- introduced by Kane-Mele for the graphene lattice model based on $p_z$ orbitals \cite{Kane_Mele}, where the orbital degrees of freedom is quenched -- is responsible for the gap opening at $\mathrm{K}$.  The kagome lattice therefore provides a model platform for studying the Kane-Mele type spin-orbit coupling and its interplay with massive Dirac fermions. Clarifying the underlying microscopic mechanisms for such spin-orbit coupling terms will provide insights in future design of topological phases from the lattice point of view.

\subsection{Interplane hopping and layer degrees of freedom of Dirac fermions in \ce{Fe3Sn2}}

\begin{figure}[h]
	\includegraphics[width = 0.85 \columnwidth]{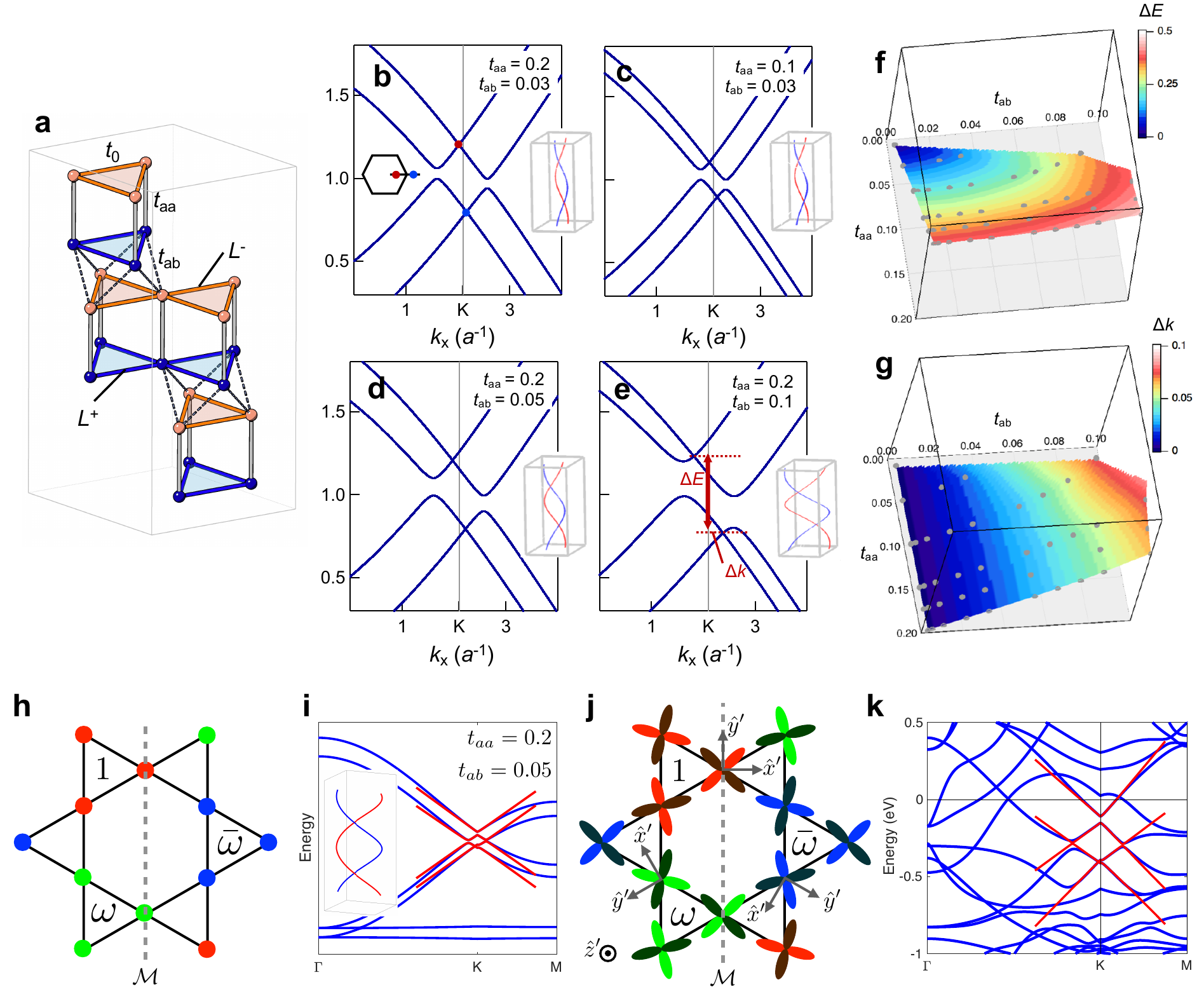}
	\caption{\label{fig-4} \textbf{Layer splitting and interplane hopping in \ce{Fe3Sn2}}
(a) Schematic of AA-BB-CC kagome lattice model with in-plane hopping $t_0$ (dark solid bonds), interplane hopping $t_{aa}$ (light solid bonds) and $t_{ab}$ (dashed bonds). (b-e) Double Dirac structure near $\mathrm{K}$ (momentum line illustrated in (b) inset) at selected $t_{aa}$ and $t_{ab}$ ($t_0=1$): (b) $t_{aa}=0.2,t_{ab}=0.03$, (c) $t_{aa}=0.1,t_{ab}=0.03$, (d) $t_{aa}=0.2,t_{ab}=0.05$, (e) $t_{aa}=0.2,t_{ab}=0.1$. The insets show the upper (red) and lower (blue) nodal lines. Here the momentum $k_x$ is expressed in the unit of $a^{-1}$, where $a$ is the in-plane hexagonal lattice constant. (f,g) Contour plots of the energy splitting $\Delta E$ (f) and momentum displacement $\Delta k$ (g) in the $t_{aa}-t_{ab}$ phase space. (h) Initial wave function on a single kagome layer used to project out double Dirac structure in AA-BB-CC model. The color illustrates the phase of wavefunction: 1 (red), $\omega=e^{i2\pi/3}$ (green) and $\bar{\omega}=e^{-i2\pi/3}$ (blue). The other partner state is obtained by the mirror $\mathcal{M}$ operation defined by the dashed grey line. (i) The $\bm{k}\cdot\bm{p}$ 4-band model (red) compared to the full 6-band AA-BB-CC model. (j) Initial basis set of a single Fe kagome layer for $\bm{k}\cdot\bm{p}$ projection of the double Dirac cones in \ce{Fe3Sn2}. The rotated local coordinate frames for Fe sites are shown with the out-of-plane $\hat{z}'$ axis. The wave function for projection is the product of the phases in (h) and the local $d_{xy}$ orbitals. The partner state on the same layer can be obtained by mirror $\mathcal{M}$. (k) \ce{Fe3Sn2} band structure (blue) compared with the projected 4-band $\bm{k}\cdot\bm{p}$ expansion (red) near K point.}
\end{figure}

Having demonstrated that both upper and lower helical nodal lines in \ce{Fe3Sn2} can be captured by quasi-2D Dirac fermions subject to a Kane-Mele type spin-orbit coupling, in the following we examine the origin of the pair of Dirac fermions in the system. Expanding from the A-B-C stacked kagome model described above, we build a tight-binding model of an AA-BB-CC stacked kagome lattice illustrated in Fig. \ref{fig-4}(a) to more accurately capture the iron sublattice of \ce{Fe3Sn2}. A fundamental rhombohedral unit cell of this model includes six atoms, forming a pair of A-A stacked kagome bilayer. This pair provides a layer degree of freedom whose role we elucidate hereafter. Aside from the in-plane nearest neighbor hopping $t_0$, we introduce two inequivalent inter-plane hopping integrals $t_{aa}$ and $t_{ab}$. $t_{aa}$ represents vertical hopping processes between aligned A-A (B-B, C-C) stacked sublattices, while $t_{ab}$ denotes the nearest neighbor hopping between layers that are rotated by $60^{\circ}$ with each other (\textit{i.e.}, through A-B, B-C and C-A stacking). 

First we find that parameter sets satisfying $t_{ab}<t_{aa}\ll t_0$ reproduces the experimental and numerical double Dirac structure in \ce{Fe3Sn2}: bands obtained from several such $t_{aa}$ and $t_{ab}$ are shown in Fig. \ref{fig-4}(b-e) and the corresponding nodal lines are shown as insets. The momentum line is highlighted in Fig. \ref{fig-4}(b) inset. We note that setting $t_{ab}>t_{aa}$ considerably deforms the Dirac bands (Supplementary Materials Fig. \ref{fig-si-tAB}) and yields band features inconsistent with either ARPES \cite{Fe3Sn2_ARPES} or the DFT spectrum shown in Fig. \ref{fig-2}(c). Hereon we focus on the evolution of the double Dirac structure with respect to $t_{aa}$ and $t_{ab}$. In Fig. \ref{fig-4}(b,d,e) a progressively increasing $t_{ab}$ displaces the nodal lines farther from K, consistent with the simpler A-B-C model (see Supplementary Materials Fig. \ref{fig-si-nl-hopping}); meanwhile the energy splitting between upper and lower nodal lines stays constant. In contrast, by varying $t_{aa}$ while keeping $t_{ab}$ constant (Fig. \ref{fig-4}(b,c)), the location of the nodal lines are unchanged while the energy splitting between upper and lower branches increases in proportion to $t_{aa}$. The respective dependence on $t_{aa}$ and $t_{ab}$ of the energy splitting $\Delta E$ and momentum displacement $\Delta k$ (both schematically illustrated in Fig. \ref{fig-4}(e)) is also clear in the contour plots of $\Delta E$ and $\Delta k$ in the $t_{aa}-t_{ab}$ phase space (Fig. \ref{fig-4}(f,g)). Further analysis of the eigenstates of the tight-binding Dirac states reveals that the upper/lower branches are predominately composed of bonding/antibonding superpositions of states residing respectively in layers $L^+$ and $L^-$ in Fig. \ref{fig-4}(a), which are connected via $t_{aa}$. 

An outstanding observation here is that $t_{aa}$ and $t_{ab}$ appear to play distinct roles to the Dirac fermions. It is instructive to adopt the following $4\times4$ $\bm{k}\cdot\bm{p}$ model in the vicinity of $\mathrm{K}$: 
\begin{equation}
\begin{split}
\mathcal{H}  = & i\hbar v_F (k_+ \sigma_- -k_- \sigma_+ ) +t_{aa} (e^{i k_z c_1} \tau_+ +e^{-i k_z c_1} \tau_-) \\
& +2 t_{ab} (e^{-i k_z c_2} \tau_+ \sigma_- + e^{i k_z c_2} \tau_- \sigma_+)
\end{split}\label{H_eff}
\end{equation} where $\bm{\sigma}$ and $\bm{\tau}$ are Pauli matrices and $v_F$ is the Dirac velocity. $c_1(c_2)$ represents the vertical distance of $t_{aa}$($t_{ab}$) hopping. $\bm{\sigma}$ represents the Dirac spinor per kagome layer (the basis wave function of $\bm{\sigma}$ within each layer is illustrated in Fig. \ref{fig-4}(h)) while $\bm{\tau}$ denotes the layer degree of freedom where $\tau_z|L^{\pm}\rangle=\pm1|L^{\pm}\rangle$. In Fig. \ref{fig-4}(i) we illustrate the $\bm{k}\cdot\bm{p}$ dispersion (red) as compared with the AA-BB-CC tight-binding model (blue) for $t_{aa}=0.2, t_{ab}=0.05$. With $t_{ab} < t_{aa}$, one can treat the last term of Eqn. \ref{H_eff} as a perturbation and project the four states to two sets; the resulting two eigenstate sub-sectors $D^{u,l}$ (upper ($u$) and lower ($l$) Dirac fermions) can be classified with $e^{i k_z c_1} \tau_+ +e^{-i k_z c_1} \tau_- = \xi^{u,l}$ ($\xi^u=1$ and $\xi^l=-1$ respectively), which are energetically split as $\xi^{u,l} t_{aa}$. Projecting the $t_{ab}$ terms into each sector, we derive a $2\times2$ effective Hamiltonian as $\mathcal{H}^{u,l}_{\text{eff}}  = i \hbar v_F (k_+ \sigma_- -k_- \sigma_+ ) +\xi^{u,l} t_{ab} (e^{i k_z c} \sigma_+ + e^{-i k_z c} \sigma_-)+ \xi^{u,l} t_{aa}$, where the Dirac points are moved to $\Delta k_x + i \Delta k_y = i \xi^{u,l} \frac{t_{ab}}{v_F} e^{-i k_z c}$ ($c=c_1+c_2$ the vertical distance between the neighboring bilayer units), giving rise to helical nodal lines (see inset of Fig. \ref{fig-4}(i)). From the $\bm{k}\cdot\bm{p}$ formulation, it is clear that $t_{ab}$ preserves an underlying sublattice (chiral) symmetry for the Dirac fermions \cite{Hatsugai}, as a result of which $t_{ab}$ terms perturb band touching points away from $\mathrm{K}$ and $\mathrm{K'}$ but cannot  generate either a gap or an energy shift to the degeneracy points. A similar $\bm{k}\cdot\bm{p}$ model can be constructed for the DFT structure as we show in Fig. \ref{fig-4}(j,k). The leading parameters are $v_F=3.8\times10^5$ m/s, $t_{aa} = -0.13$ eV, $t_{ab} =  -0.0028$ eV (also see Methods), suggestive of a layer split nature of the two copies of Dirac fermions in \ce{Fe3Sn2} \cite{Fe3Sn2_ARPES}. The layer origin of the upper and lower Dirac states is also consistent with the similar Berry curvature structure they exhibit as illustrated in Fig. \ref{fig-3}(d).

\begin{table}[t]
  \centering
  \caption{The Dirac wave function $k_z$-averaged ($0 \leq k_z < 2\pi/c$) density distribution for Fe and Sn atomic orbitals, at the projected (2D) K point.  Sn$^{(s)}$ (Sn$^{(k)}$) denotes the set of Sn atoms in spacer unit (kagome layer). The atomic orbitals are defined with the rotated local coordinate as shown in Fig. 4(j). The density is represented by the percentage projected in atomic orbitals.} 
  \label{tab:dirac_cone_wf}
  \begin{tabular}{ccccccccccc}
  \hline
 & Fe $d_{xy}$ & Fe $d_{x^2-y^2}$ & Fe $d_{xz}$ & Fe $d_{yz}$ &  Fe $d_{z^2}$ & Fe $s$ & Sn$^{(s)}$ $p_z$ & Sn$^{(s)}$ $p_{x/y}$ & Sn$^{(k)}$ $p$ & Sn $s$ \\
  \hline
Lower Dirac cone & 33.0 & 23.8 & 10.9 & 9.9 & 9.1 & 1.7 & 0.4 & 9.2 & 1.5 & 0.5\\
Upper Dirac cone & 49.4 & 9.6 & 16.6  & 15.0 & 4.3 & 1.3 & 0.2 & 1.8 & 1.6 & 0.3\\
  \hline
  \end{tabular}
\end{table}

We additionally analyzed the orbital characters of the Wannier function for the upper and lower Dirac fermions as summarized in Table 1. The predominant in-plane nature of both upper and lower Dirac fermions dictates reduced strength of $t_{aa}$ and $t_{ab}$ as compared to $t_0$ (we may estimate $t_0\sim0.8$ eV from $v_F$ via $t_0\simeq\sqrt{3}\hbar v_F/a$, where $a$ is the in-plane lattice constant), which further leads to the characteristic double Dirac structure in \ce{Fe3Sn2} (see Supplementary Materials for a more detailed description of the orbital-decomposed hopping pathways). Moreover, as can be inferred from Eqn. \ref{H_eff}, $t_{aa}$ and $t_{ab}$ are decoupled from the $k_z$-evolution of the nodal line energies, implying that higher order hopping terms are required to grant the Dirac fermions a considerable $k_z$ dispersion. This is consistent with the suppressed $k_z$-dispersion and quasi-two-dimensionality of the bulk Dirac fermions suggested experimentally in \ce{Fe3Sn2} \cite{Fe3Sn2_ARPES,Fe3Sn2_dHvA}. We note that including an asymmetry to account for the breathing nature of the kagome lattices in \ce{Fe3Sn2} does not considerably alter the scenarios presented here (see Fig. \ref{fig-si-breathing}). Through the above minimal AA-BB-CC model and $\bm{k}\cdot\bm{p}$ expansion we establish \ce{Fe3Sn2} as an illustrative example of how rich stacking patterns and associated interplane coupling manifest in quasi-two-dimensional electronic materials. In Fig. \ref{fig-si-graphite} we illustrate that similar results can be obtained for a model of AA-BB-CC stacked honeycomb layers. 

\subsection{Discussion}

In summary, from a theoretical perspective, we have established \ce{Fe3Sn2} as a host of ferromagnetic helical nodal lines derived from a kagome network of iron. The peculiar presence of mixed A-A and A-B stacking patterns of kagome lattices in \ce{Fe3Sn2} causes the formation of helical nodal lines, gives rise to the layer splitting between upper and lower branches, and suppresses the $k_z$-dispersion of these nodal lines. With an out-of-plane ferromagnetic order, the two sets of helical nodal lines are gapped out by spin-orbit coupling and serve as strong source of Berry curvatures. Gradually rotating the ferromagnetic moments from out-of-plane to in-plane orientations one may partially close the Dirac mass gap at discrete points and realize pairs of Weyl nodes located along the original helical nodal lines. In view of the soft ferromagnetic nature of the system we may anticipate novel electronic states at domain walls \cite{Nomura_DW}; one especially exciting avenue lies in the skyrmion bubble structures observed earlier in \ce{Fe3Sn2} at room temperature \cite{Fe3Sn2_skyrmion}, where the real space topological spin textures may entangle with the nodal lines and give rise to novel electronic responses \cite{Graphene_skyrmion}. 

A direct experimental observation of the drumhead surface states in \ce{Fe3Sn2} has remained elusive due in part to the weak interplane coupling of the $t_{ab}$ form and a resulting limited radial size of the helical nodal lines. Further engineering of the interplane hopping and spin-orbit coupling in related intermetallic compounds that host trigonal stacking of kagome lattices may lead to experimentally detectable drumhead surface states \cite{Graphite_ARPES}.  We note that a recent tight-binding study of a $d_{z^2}$ type orbital on an A-B-C kagome lattice with enhanced out-of-plane hopping assisted by interlayer Sn atoms serves as a minimal model to generate vertical nodal rings and the ferromagnetic Weyl semimetallic phase in \ce{Co3Sn2S2} \cite{Co3Sn2S2_Nomura}. This comparison suggests that with rational orbital engineering,  intermetallic compound-based kagome lattices may provide a full spectrum of topological phases ranging from the 3D quantum anomalous Hall insulating phase \cite{Fe3Sn2_ARPES,FeSn,MTI_layer} to the ferromagnetic Weyl semimetallic phase \cite{Co3Sn2S2}; driving the topological phase transition between the two classes of phases are of extreme theoretical and experimental interest. 

The implications of our study are beyond electronic structures of metallic systems. As trigonal stacking and rhombohedral symmetry are ubiquitous in naturally occuring kagome lattice materials \cite{Cs2LiMn3F12,herbertsmithite,kagome_Dirac,Na2Ti3Cl8}, including, for instance, the spin liquid hosting herbertsmithite \cite{herbertsmithite}, we expect that the helical nodal lines discussed here may be relevant not only in the electronic sector, but also in the magnonic or spinonic sectors \cite{DiracSL}, in the context of considerable inter-plane coupling. In view of the close resemblance of the kagome lattice with the honeycomb lattice \cite{Graphite_JETP,ABC_Graphite_2016,CrCl3_magnon}, the above picture could be relevant, for example, in the Dirac-like Majorana fermionic spectrum in $\alpha-$RuCl$_3$ where the Ru honeycomb layers are A-B-C stacked \cite{RuCl3}.

\section{Methods}
\subsection{{\bf Density functional theory electronic structure calculations}}

To compute the electronic and related properties of Fe$_3$Sn$_2$ we carry out the Density Functional Theory (DFT) calculations by using the full-potential local-orbital (FPLO) code~\cite{FPLO}, version 18.00-52. The exchange-correlation energy functional used is based on the parameterization of Perdew, Burke, and Ernzerhof (PBE-96)~\cite{FPLO_PBE} within the generalized gradient approximation. A linear tetrahedron method with $ 15 \times 15 \times 15$ subdivisions in the full Brillouin zone was used for the momentum space integrations. The lattice parameters used in the calculation are $a=5.3307\mathrm{\AA}$ and $c=19.7968\mathrm{\AA}$ \cite{Fe3Sn2_MP}. We consider the ground state in the ferromagnetic state and converge the self-consistent calculations within the scalar relativistic mode, and (four-component) fully relativistic mode of FPLO, with a self-consistent spin density better than $10^{-6}$. The total magnetic moments per unit cell (\ce{Fe6Sn4}) for the converged ground states are 12.21 $\mu_B$ in scalar relativistic mode, 12.64 $\mu_B$ and 12.65 $\mu_B$ respectively for the fully relativistic mode with ferromagnetic state along [001] and [100] orientations. We note that the latter two values
include orbital magnetic moments.

To carry out further analysis of the electronic structure, we derive the Wannier tight-binding Hamiltonian by projecting the Bloch states onto atomic orbital-like Wannier functions using the PYFPLO module of the FPLO package~\cite{FPLO}. These localized Wannier basis states include Fe $4s$, $3d$, orbitals and Sn $5s$, $5p$ orbitals. The Wannier model is converged with a $8 \times 8 \times 8$ grid sampling in the Brillouin zone.  These derived Wannier Hamiltonians are then used to investigate the nodal Dirac structure and associated topological properties such as Berry curvatures~\cite{Berry_numerical_method}.

\subsection{{\bf \ce{Fe3Sn2} $\bm{k}\cdot\bm{p}$ expansion and calculation of drumhead surface states}}

Here we give a more detailed $\bm{k}\cdot\bm{p}$ expansion for \ce{Fe3Sn2} electronic structure near the double Dirac cones, as performed for the AA-BB-CC model in Eq. \ref{H_eff}. The numerical projection is based on the Wannier construction (see main text and supplementary materials). The four-band effective Hamiltonian can be summarized as:
\begin{equation}
\begin{split}
\mathcal{H}  = & i\hbar v_F (k_+ \sigma_- -k_- \sigma_+ ) +E_0 + ( t_{aa} e^{-i k_z c_1} \tau_- + t_{ab} e^{-i k_z c_2} \tau_+ \sigma_- + {\rm h.c.}) \\
& + (t_1 e^{-i k_z (c_1+c_2)} \sigma_- + t_2 e^{-i k_z (2c_1+c_2)} \sigma_- \tau_- + t_3 e^{-i k_z (c_1+2c_2)} \sigma_+ \tau_+ + t_4 e^{-i 2 k_z (c_1+c_2)} \sigma_+ + {\rm h.c.})
\end{split}\label{Fe3Sn2_kp_eff}
\end{equation} where $\hbar v_F = 2.52\mathrm{eV}\cdot\mathrm{\AA}$ (corresponding to $v_F=3.8\times10^5$ m/s), $E_0 = -0.26$ eV, $t_{aa} = -0.13$ eV, $t_{ab} =  -0.0028$ eV, $t_{1} = 0.0123$ eV, $t_{2} = 0.0059$ eV, $t_{3} = -0.0339$ eV and $t_{4} = -0.008$ eV. This effective model captures the helical nodal line structure for both upper and lower cones. Starting from this, we consider a finite thin-film slab geometry to shed light on the surface states associated with the helical nodes in \ce{Fe3Sn2}. We found the drumhead surface states near K points as illustrated in Fig. \ref{fig-2}(k) in the surface spectral function for the lower Dirac cone (similar drumhead surface states can be obtained for the upper Dirac cone). 

\section{Acknowledgement}
We are grateful to B. Lian, J.-S. You, and T. Kurumaji for fruitful discussions. This work was funded, in part, by the Gordon and Betty Moore Foundation EPiQS Initiative, Grant No. GBMF9070 to J.G.C. and NSF grant DMR-1554891. L.Y. and E.K. acknowledge support by the STC Center for Integrated Quantum Materials, NSF grant number DMR-1231319. L.Y. acknowledges support from the Heising-Simons Foundation. S.F. is supported by a Rutgers Center for Material Theory Distinguished Postdoctoral Fellowship. M.P.G. acknowledges the equipment grant supported by Alexander von Humboldt Foundation, Germany. J.v.d.B. acknowledges financial support from the German Research Foundation (Deutsche Forschungsgemeinschaft, DFG) via SFB1143 Project No. A5 and under German Excellence Strategy through the W\"urzburg-Dresden Cluster of Excellence on Complexity and Topology in Quantum Matter ct.qmat (EXC 2147, Project No. 390858490). The computations in this paper were run on the FASRC Cannon cluster supported by the FAS Division of Science Research Computing Group at Harvard University and at the computer clusters at IFW Dresden, Germany. S.F., M.P.G. and M.R. acknowledge the technical assistance from U. Nitzsche for the latter. J.L. acknowledges the support from the Hong Kong Research Grants Council (26302118, 16305019 and N\textunderscore HKUST626/18).\\

%\section{Author contributions}
%S.F. and L.Y. performed the theoretical modelings with inputs from M.G.K., L.F., J.L.,and R.C.. S.F. and M.P.G. performed the DFT calculations with J.L., M.R., J.v.d.B, and E.K. supporting.  L.Y. and S.F. wrote the manuscript with contributions from all authors. J.G.C. supervised the project. \\

%\section{Data availability}
%The data that support the findings of this study are available from the corresponding author on reasonable request.\\

%\section{Code availability}
%The codes used to support the findings in this study are available from the corresponding author on reasonable request.\\

%\section{Competing interests}
%The authors declare no competing interests. \\

\newpage

\clearpage
\newpage

%\title{Supplementary Information\\Ferromagnetic helical nodal line and Kane-Mele spin-orbit coupling in kagome metal \ce{Fe3Sn2}}
%\maketitle

\begin{center}
    \textbf{Supplementary Information\\Ferromagnetic helical nodal line and Kane-Mele spin-orbit\\coupling in kagome metal \ce{Fe3Sn2}}
\end{center}

\clearpage
\newpage

\section{\NoCaseChange{Evolution of helical nodal line with interplane hopping}}

\begin{figure}[h]
	\includegraphics[width = \columnwidth]{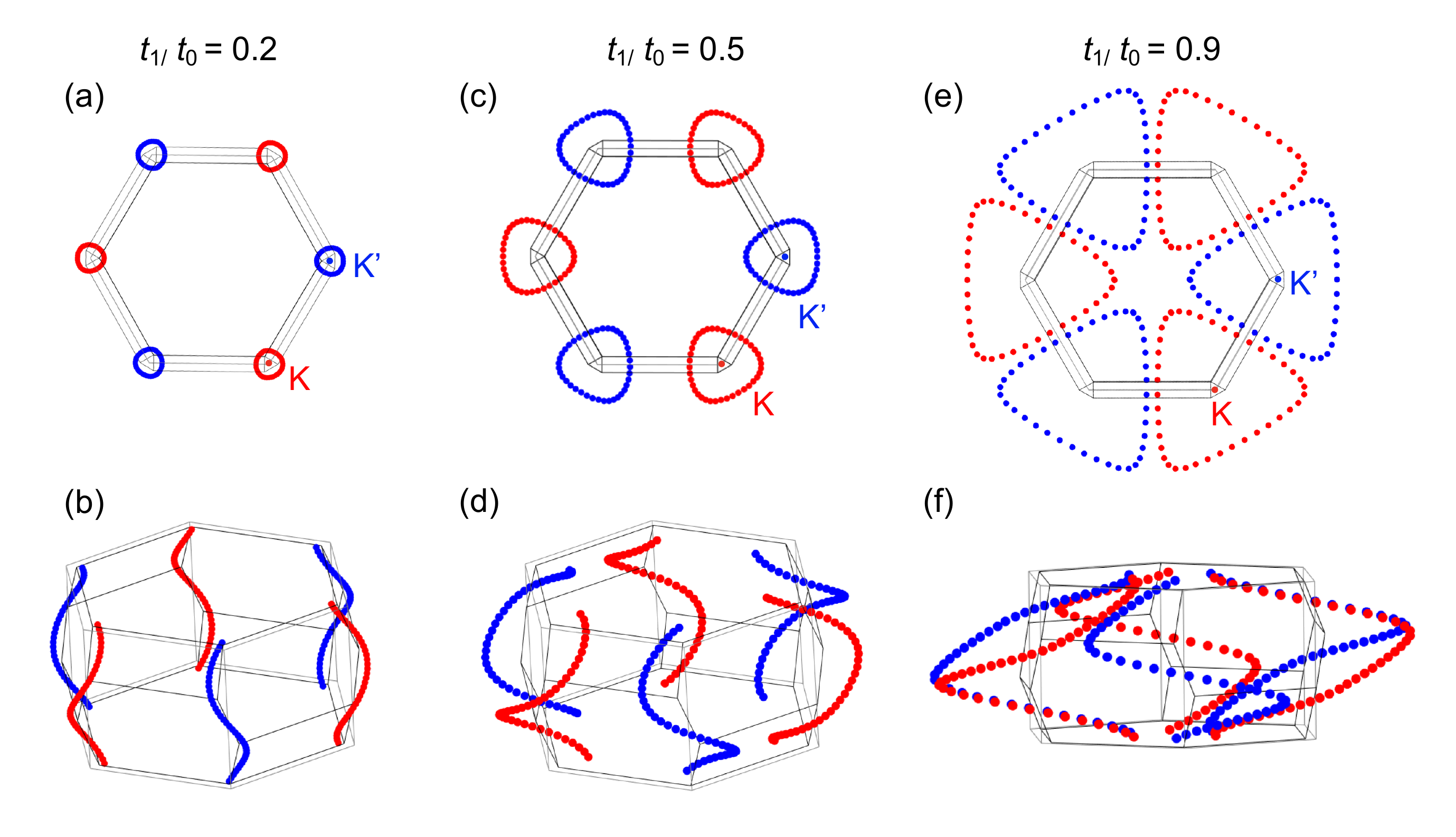}
	\caption{\label{fig-si-nl-hopping} The nodal line structure of the A-B-C kagome lattice model. (a) Top and (b) three-dimensional view of nodal lines with $t_1/t_0=0.2$; (c) top and (d) three-dimensional view with $t_1/t_0=0.5$; 	(e) top and (f) three-dimensional view with $t_1/t_0=0.9$. The red and blue nodal lines represent those originated from the 2D Dirac fermions at $\mathrm{K}$ and $\mathrm{K}'$, respectively. }
\end{figure}

In Fig. \ref{fig-si-nl-hopping} we examine the evolution of the helical nodal lines with a gradually increasing out-of-plane hopping $t_1$, where Fig. \ref{fig-si-nl-hopping}(a,b), (c,d), (e,f) show the top and 3D views with $\dfrac{t_1}{t_0}=0.2,0.5,0.9$, respectively. We find that in the limit $\dfrac{t_1}{t_0}\ll1$, the trajectory of the nodal line (here the origin is set at $\mathrm{K}$ and $\mathrm{K}$' of the hexagonal Brillouin zone) may be described by the following polar coordinate $(r,\phi)$ at each $k_z$:
\begin{equation}\label{NNNL}
\begin{array}{c}
r = \dfrac{2.37}{a}(\dfrac{t_1}{t_0})\\
\phi=\eta(k_zc)\\
\end{array}
\end{equation}
where the chirality $\eta$ takes $1(-1)$ at $\mathrm{K}$ ($\mathrm{K}'$), respectively. Here we use $a$ to denote the kagome lattice constant and $c$ the vertical interlayer distance. The radial displacement from $\mathrm{K}/\mathrm{K}'$ is roughly proportional to $\dfrac{t_1}{t_0}$ and with a growing $\dfrac{t_1}{t_0}$ a trigonal warping of the nodal line trajectories becomes apparent (see Fig. \ref{fig-si-nl-hopping}(c,e)). Nevertheless the existence of the nodal lines remains robust as long as the blue and red nodal lines do not touch.

\section{\NoCaseChange{Drumhead surface states in kagome stacks}}

\begin{figure}[h]
	\includegraphics[width = 0.95\columnwidth]{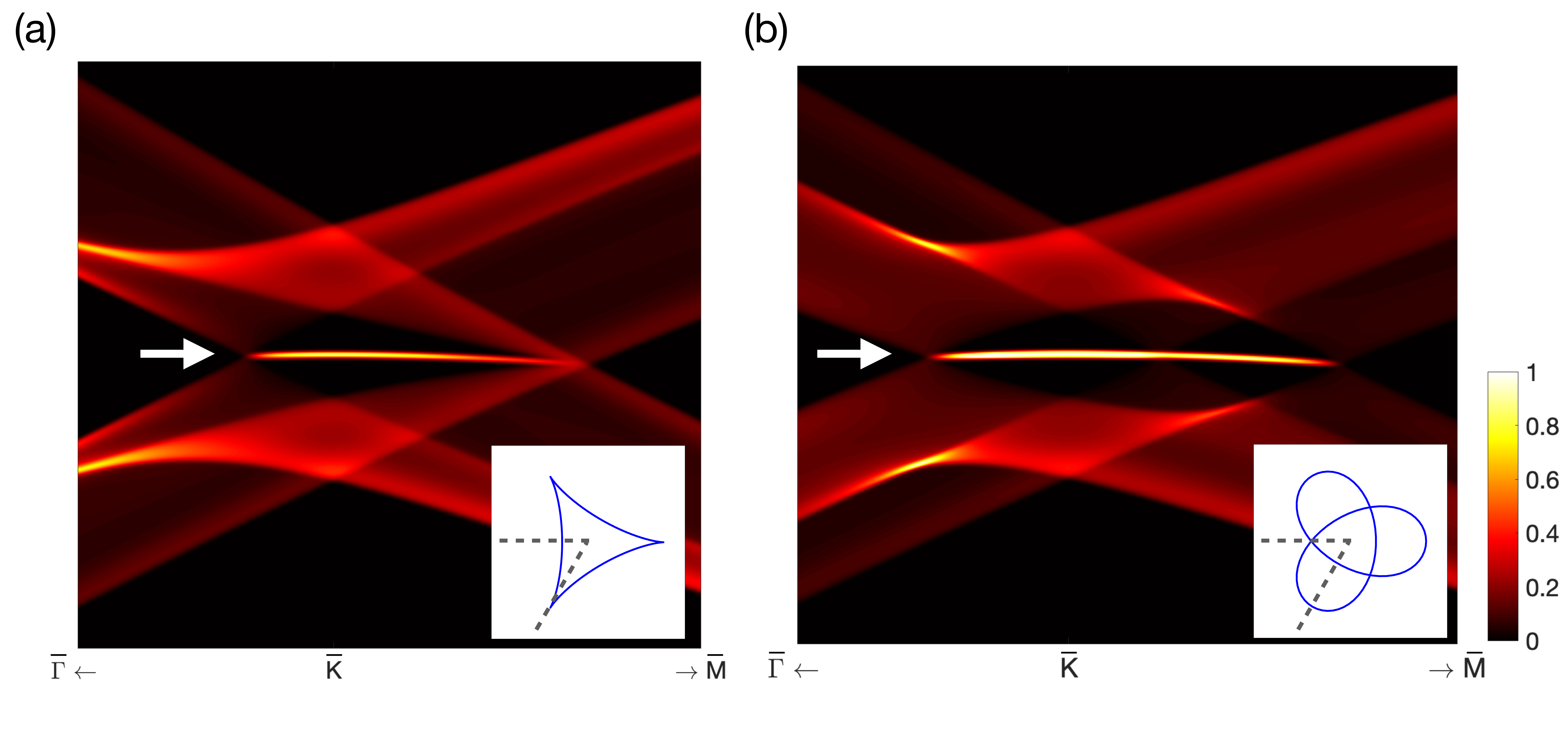}
	\caption{\label{fig-si-ABC-DSS} The helical nodal line structure in the ABC-stacked kagome crystal, and the associated drumhead surface states (indicated by white arrows) in the surface spectral function for (a) $t_1 > t_2$ with $t_1=0.1$ and $t_2=0.05$ (b) $t_2 > t_1$ with $t_1=0.05$ and $t_2=0.1$. The in-plane coupling $t_0$ is taken as unity. Insets show the projection of the nodal lines in the $k_x-k_y$ plane as blue solid lines. The dashed gray lines depict the $\mathrm{\Gamma}-\mathrm{K}-\mathrm{M}$ momentum cuts. The surface spectral function integrate the weights for the first five kagome layers.}
\end{figure}

Here we first construct a simple ABC stacked kagome lattice model to illustrate the drumhead head surface states associated with the helical nodal lines. In an ABC-stacked kagome slab crystal, with s-wave orbitals on kagome sites, the Hamiltonian $H(\vec{k}_{\parallel})$ can be derived as
\begin{equation}
    H(\vec{k}_{\parallel})=\sum_{i,j}[\delta_{i,j} H_0(\vec{k}_{\parallel})+\delta_{j,i+1} H_1(\vec{k}_{\parallel})+\delta_{j,i-1} H_{-1}(\vec{k}_{\parallel})+\delta_{j,i+2} H_2(\vec{k}_{\parallel})+\delta_{j,i-2} H_{-2}(\vec{k}_{\parallel})]
\end{equation} where $i,j$ are the layer unit index within the slab, and 2D momentum $\vec{k}_{\parallel}=(k_x,k_y)$. $H_0$ stands for intraplane hopping while $H_{\pm1}$ and $H_{\pm2}$ feature the nearest and next-nearest hopping, respectively:
\begin{equation}
\begin{split}
     H_0(\vec{k}_{\parallel})= t_0 & \begin{bmatrix}
    0 & e^{-i \vec{k}_{\parallel} \cdot \vec{\delta_1}}+e^{i \vec{k}_{\parallel} \cdot \vec{\delta_1}} & e^{-i \vec{k}_{\parallel} \cdot \vec{\delta_2}}+e^{i \vec{k}_{\parallel} \cdot \vec{\delta_2}} \\
    e^{-i \vec{k}_{\parallel} \cdot \vec{\delta_1}}+e^{i \vec{k}_{\parallel} \cdot \vec{\delta_1}} & 0 & e^{-i \vec{k}_{\parallel} \cdot \vec{\delta_3}}+e^{i \vec{k}_{\parallel} \cdot \vec{\delta_3}} \\
    e^{-i \vec{k}_{\parallel} \cdot \vec{\delta_2}}+e^{i \vec{k}_{\parallel} \cdot \vec{\delta_2}} & e^{-i \vec{k}_{\parallel} \cdot \vec{\delta_3}}+e^{i \vec{k}_{\parallel} \cdot \vec{\delta_3}} & 0 \\
    \end{bmatrix}, \\
    & H_1(\vec{k}_{\parallel})=t_1 \begin{bmatrix}
    0 & e^{-i \vec{k}_{\parallel} \cdot \vec{\delta_{32}}} & e^{-i \vec{k}_{\parallel} \cdot \vec{\delta_{13}}} \\
    e^{-i \vec{k}_{\parallel} \cdot \vec{\delta_{32}}} & 0 & e^{-i \vec{k}_{\parallel} \cdot \vec{\delta_{21}}} \\
    e^{-i \vec{k}_{\parallel} \cdot \vec{\delta_{13}}} & e^{-i \vec{k}_{\parallel} \cdot \vec{\delta_{21}}} & 0 \\
    \end{bmatrix}, \\
    & H_2(\vec{k}_{\parallel})=t_2 \begin{bmatrix}
    0 & e^{-i \vec{k}_{\parallel} \cdot \vec{\delta_{23}}} & e^{-i \vec{k}_{\parallel} \cdot \vec{\delta_{31}}} \\
    e^{-i \vec{k}_{\parallel} \cdot \vec{\delta_{23}}} & 0 & e^{-i \vec{k}_{\parallel} \cdot \vec{\delta_{12}}} \\
    e^{-i \vec{k}_{\parallel} \cdot \vec{\delta_{31}}} & e^{-i \vec{k}_{\parallel} \cdot \vec{\delta_{12}}} & 0 \\
    \end{bmatrix}
    \end{split}
\end{equation} and $H_{-i}(\vec{k}_{\parallel}) = H_i^{\dagger}(\vec{k}_{\parallel})$. Here $t_0$ is the intra-layer nearest neighbor coupling between kagome sites, $t_1 (t_2)$ the dominant inter-layer coupling to the first (second) nearest layer in the out-of-plane direction. The vectors used are $\vec{\delta_1}=a\hat{x}/2$, $\vec{\delta_2}=a(-\hat{x}/4-\sqrt{3}\hat{y}/4)$,  $\vec{\delta_3}=a(-\hat{x}/4+\sqrt{3}\hat{y}/4)$ and $\vec{\delta_{ij}}=(\vec{\delta_i}-\vec{\delta_j})/3$. The corresponding bulk Hamiltonian takes the following form:
% H_{-1}(\vec{k})e^{i k_z c}+ H_{-2}(\vec{k})e^{i2 k_z c}
\begin{equation}
    H_{\rm bulk}(\vec{k}_{\parallel},k_z)= H_0(\vec{k}_{\parallel})+ (H_1(\vec{k}_{\parallel})e^{-i k_z c}+H_2(\vec{k}_{\parallel})e^{-i2 k_z c}+ {\rm h.c.})
\end{equation}

\begin{figure}[t]
	\includegraphics[width = 0.95\columnwidth]{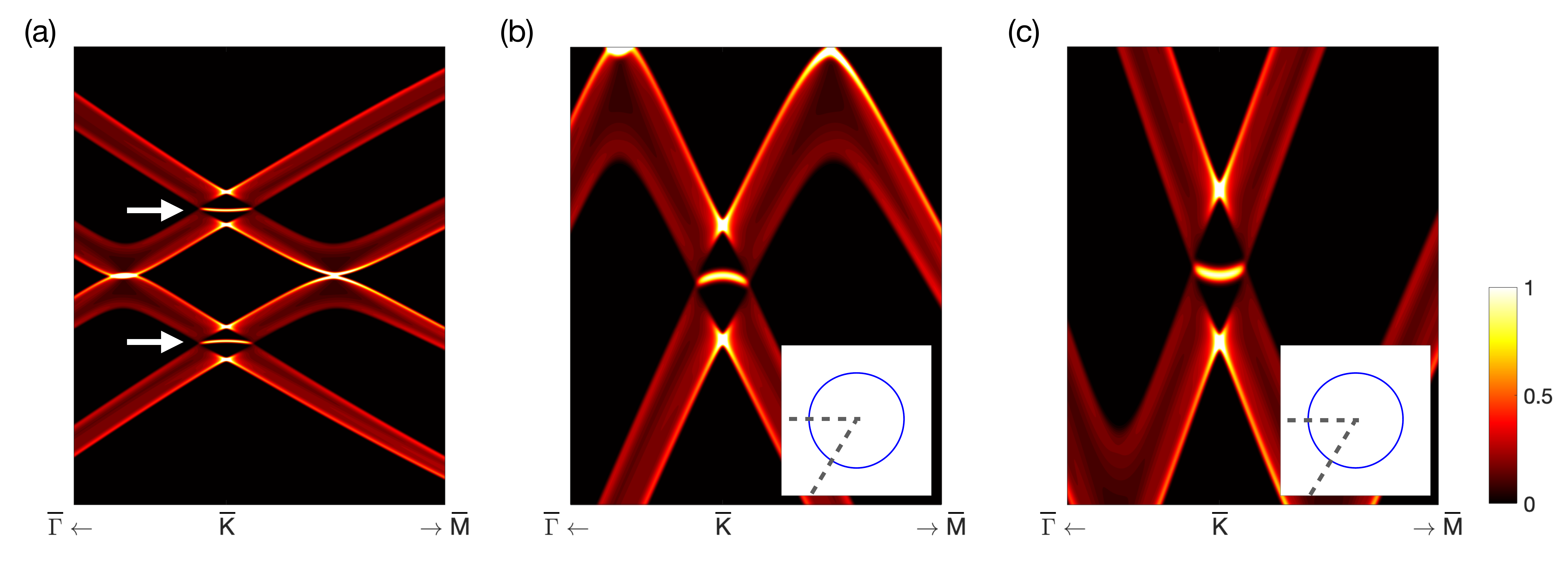}
	\caption{\label{fig-si-AABBCC-DSS} The helical nodal line structure in the AABBCC-stacked kagome crystal, and the associated drumhead surface states (indicated by white arrows) in the surface spectral function. The model has $t_{aa}=0.2$, $t_{ab}=0.05$ with the in-plane coupling $t_0$ taken as unity. (a) The surface spectral function associated with the double cone structure. (b) The surface spectral function near the lower cone with the inset showing the projection of the nodal line in the $k_x-k_y$ plane. (c) Similar analysis for the upper cone. Apart from the double cone structure, the drumhead surface states associated with each cone resemble the simpler ABC stacked model previously. More intricate nodal structure and surface states can be obtained with further interlayer couplings.}
\end{figure}

In Fig. \ref{fig-si-ABC-DSS}, we show that by introducing multiple out-of-plane hopping terms ($t_1,t_2$) we are capable of reproducing the three-fold hypotrochoid patterns of the helical nodal lines. The helical nodal line structures in the bulk and the associated drumhead surface states in the surface spectral function are shown in Fig. \ref{fig-si-ABC-DSS}(a) for $t_1 > t_2$ and in Fig. \ref{fig-si-ABC-DSS}(b) for $t_2 > t_1$, respectively. These nodal lines are found to be protected by the combined $\mathcal{P}\mathcal{T}$ symmetry which quantizes the Berry phase in a loop around the helical nodal line to integer multiples of $\pi$. Despite that the real crystal structure of \ce{Fe3Sn2} contains kagome bilayers, we note that this simplified ABC model captures the essential features of the drumhead surface states derived from a $\bm{k}\cdot\bm{p}$ approach as adopted in the main text. We have also verified that additionally taking into account the AA-BB-CC structure does not appear to alter the features of the nodal lines or drumhead surface states (illustrated in Fig. \ref{fig-si-AABBCC-DSS}, only showing the case with nearest layer hopping for clarity).
%The relative sizes of couplings $t_i$ determines the number of winding around the $k_z$ axis from the spiral nodal line.

%Near the projected K point, the meandering Dirac nodal line structure is shown in Fig. . $t_1/t_2$ controls the times of winding around K axis with $0\leq k_z < 2\pi/c$. Later, we will derive the effective k.p expansion which precisely captures this meandering nodal line structure. 

%When considering a slab geometry from the above crystal, with a finite number of stacked kagome layers, the Hamiltonian can be derived similarly 

\newpage
\section{\NoCaseChange{AA-BB-CC kagome and honeycomb lattice model}}

\begin{figure}[h]
	\includegraphics[width = 0.95\columnwidth]{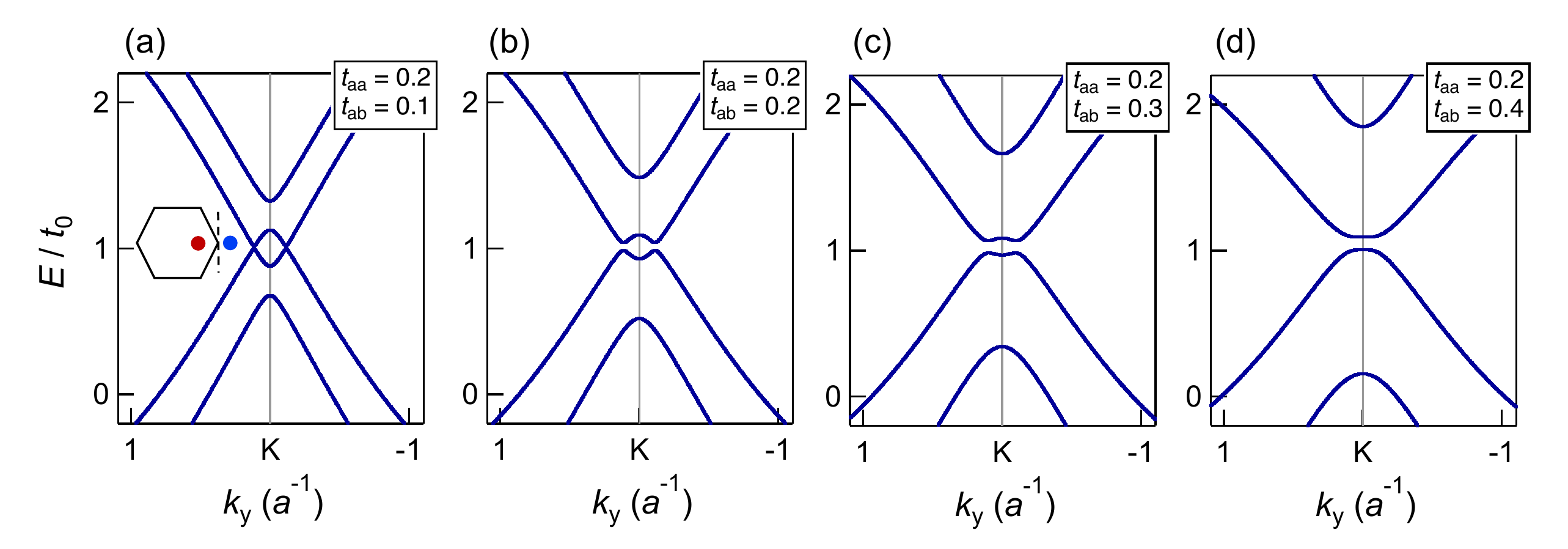}
	\caption{\label{fig-si-tAB} Evolution of the double Dirac structure with $t_{ab}$. Here $t_0$ is kept at unity and $t_{aa}$ at 0.2 while $t_{ab}=0.1$ (a), 0.2 (b), 0.3 (b) and 0.4 (d). Inset of (a) shows the momentum cut with respect to the hexagonal Brillouin zone, where the nodal line locations are schematically illustrated as red and blue dots.}
\end{figure}

\begin{figure}[h]
	\includegraphics[width = 0.8\columnwidth]{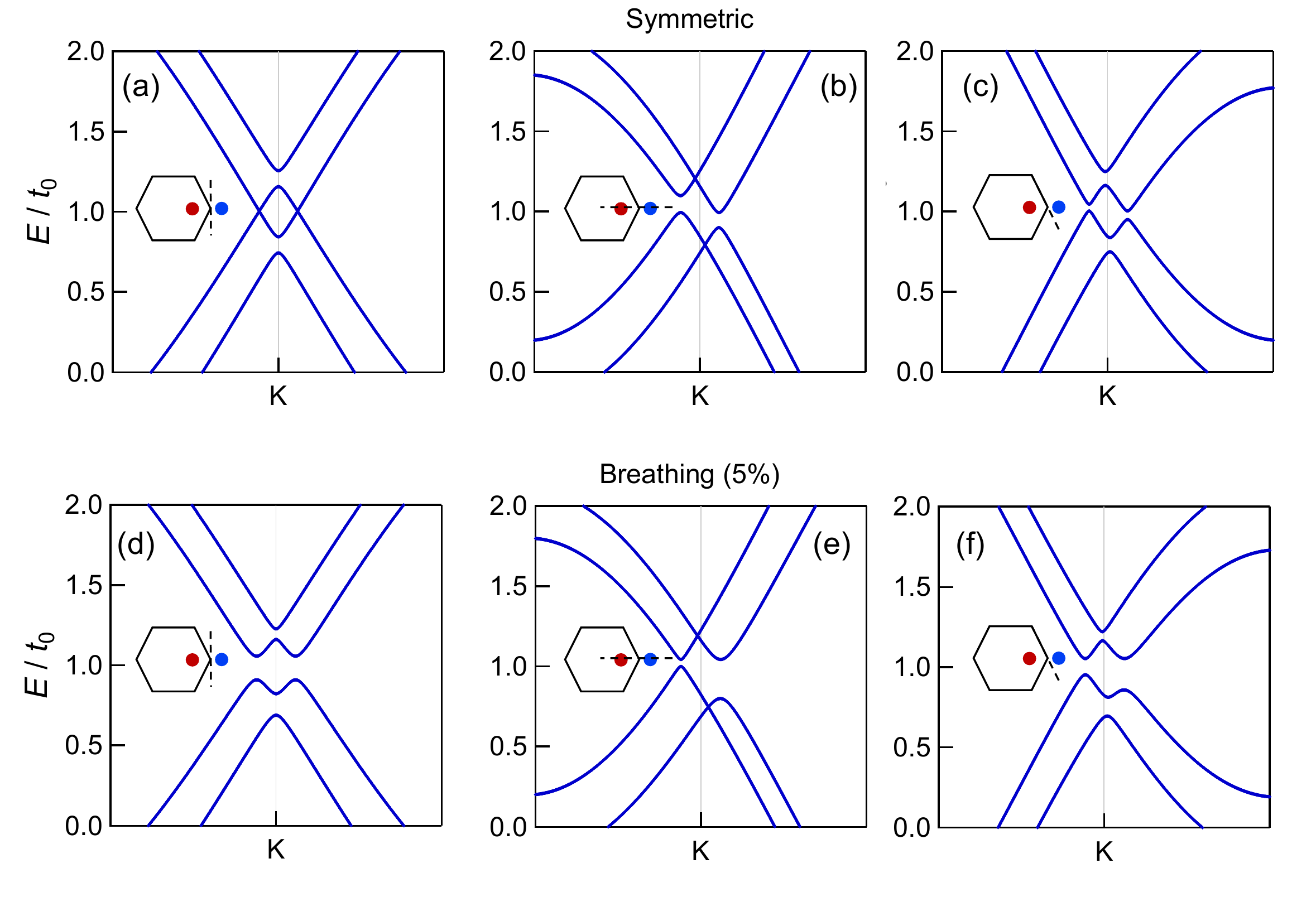}
	\caption{\label{fig-si-breathing} Modification of the double Dirac structure with inclusion of breathing nature of the kagome lattice at $k_z=0$. (a-c) illustrates the double Dirac structure with non-breathing kagome lattice while (d-f) illustrate that with the inclusion of a small breathing distortion of the kagome lattice in each layer. Two inequivalent in-plane hopping with strengths 1 and 0.95 are incorporated in this model. The momentum cuts for each panel is schematically illustrated as dashed lines in insets and the location of upper and lower Dirac crossing points are marked by red and blue dots.}
\end{figure}

In this section we provide additional results from the AA-BB-CC kagome model and also describe a closely related AA-BB-CC honeycomb lattice model. In Fig. \ref{fig-si-tAB} we explore the variation of the band structure near $\mathrm{K}$ at $k_z=0$ with an increasing strength of $t_{ab}$. In Fig. \ref{fig-si-tAB}(a,b) $t_{aa}>t_{ab}$ while in Fig. \ref{fig-si-tAB}(c,d) $t_{aa}<t_{ab}$. The vertically split double Dirac structure gradually evolves to resemble a quadratic band touching with increasing $t_{ab}$. We note that the two limiting cases ($t_{aa}\ll t_{ab}$, $t_{ab}\ll t_{aa}$) resemble the A-A stacked and A-B stacked bilayer graphene near $\mathrm{K}$ and zero energy, respectively \cite{bilayer_graphene}.

The breathing kagome lattice structure of \ce{Fe3Sn2} inspires us to explore the introduction of a small asymmetry to the double Dirac helices. By contrasting the symmetric (Fig. \ref{fig-si-breathing}(a-c)) and the asymmetric (Fig. \ref{fig-si-breathing}(d-f)) models we note that the upper and lower band crossings near $\mathrm{K}$ and their helical nature is largely unaffected. The primary effect of the asymmetry within the kagome plane beyond a threshold strength is the opening of a full gap at where the upper and lower Dirac branches meet. 

\begin{figure}[h]
	\includegraphics[width = 0.85\columnwidth]{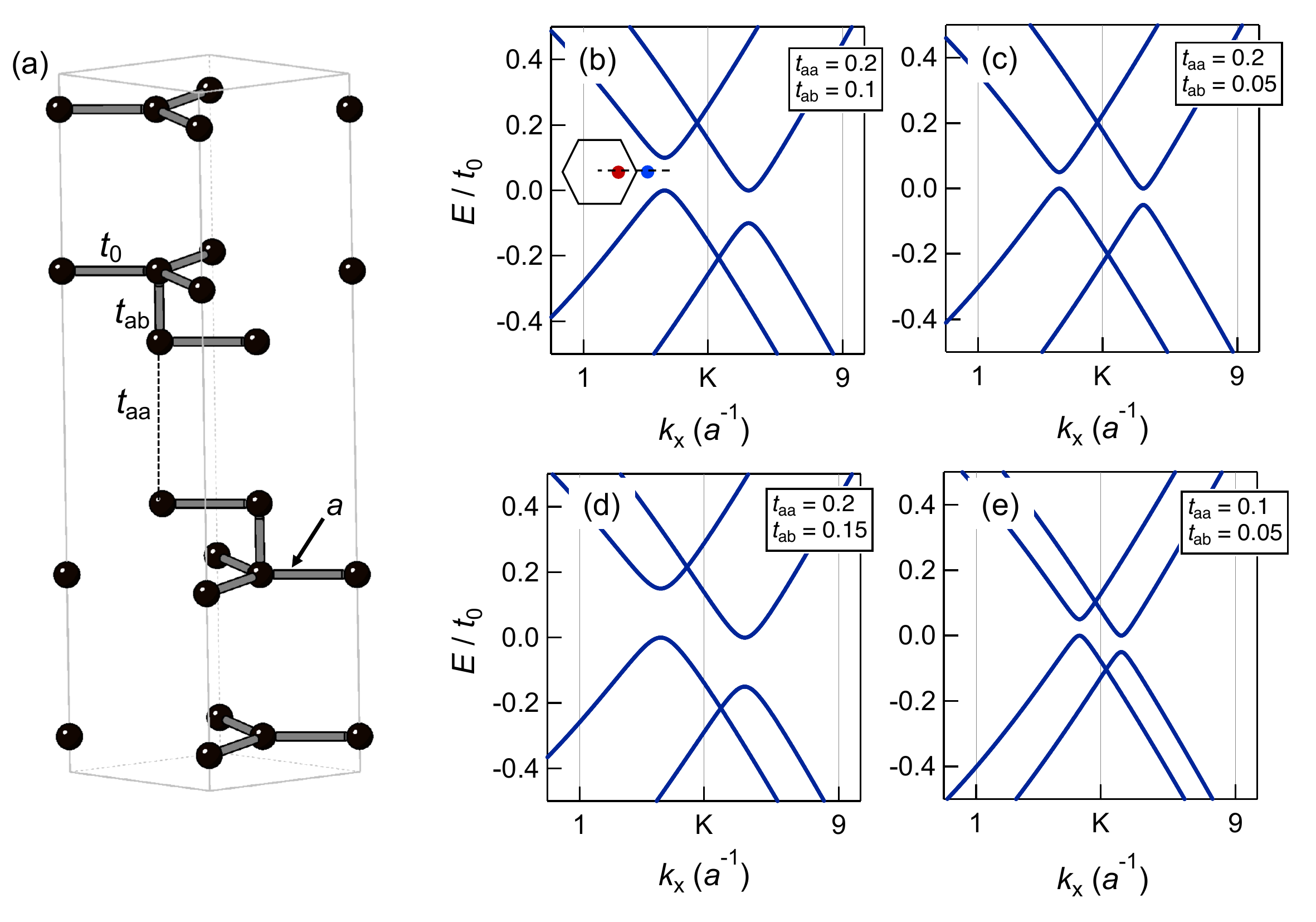}
	\caption{\label{fig-si-graphite} Results of tight-binding model of an AA-BB-CC stacked honeycomb lattice. (a) Schematic of the AA-BB-CC honeycomb lattice model with $t_0,t_{aa},t_{ab}$ highlighted. Here $a$ represents the in-plane nearest neighbor bond length. (b-e) show double Dirac structures of honeycomb lattice model with (b) $t_{aa}=0.2,t_{ab}=0.1$, (c) $t_{aa}=0.2,t_{ab}=0.05$, (d) $t_{aa}=0.2,t_{ab}=0.15$, (e) $t_{aa}=0.1,t_{ab}=0.05$. Inset of (b) illustrates the momentum cut with respect to the in-plane hexagonal Brillouin zone.}
\end{figure}

In Fig. \ref{fig-si-graphite} we show an AA-BB-CC honeycomb model which yields very similar results with the kagome model presented in the main text: $t_{aa}$ controls the splitting between upper and lower Dirac states and $t_{ab}$ displaces the nodal lines from high symmetry $\mathrm{K}$ and $\mathrm{K'}$ of the hexagonal Brillouin zone. %This implies that the notion of spiral Dirac nodal lines presented in this work is generic for 3D stacking of lattice-driven Dirac fermions.
\FloatBarrier

\section{\NoCaseChange{DFT electronic structure calculations, Wannier tight-binding Hamiltonian and applications}}

\subsection{Wannier tight-binding Hamiltonians}

\begin{figure}[h]
	\includegraphics[width = 0.6\columnwidth]{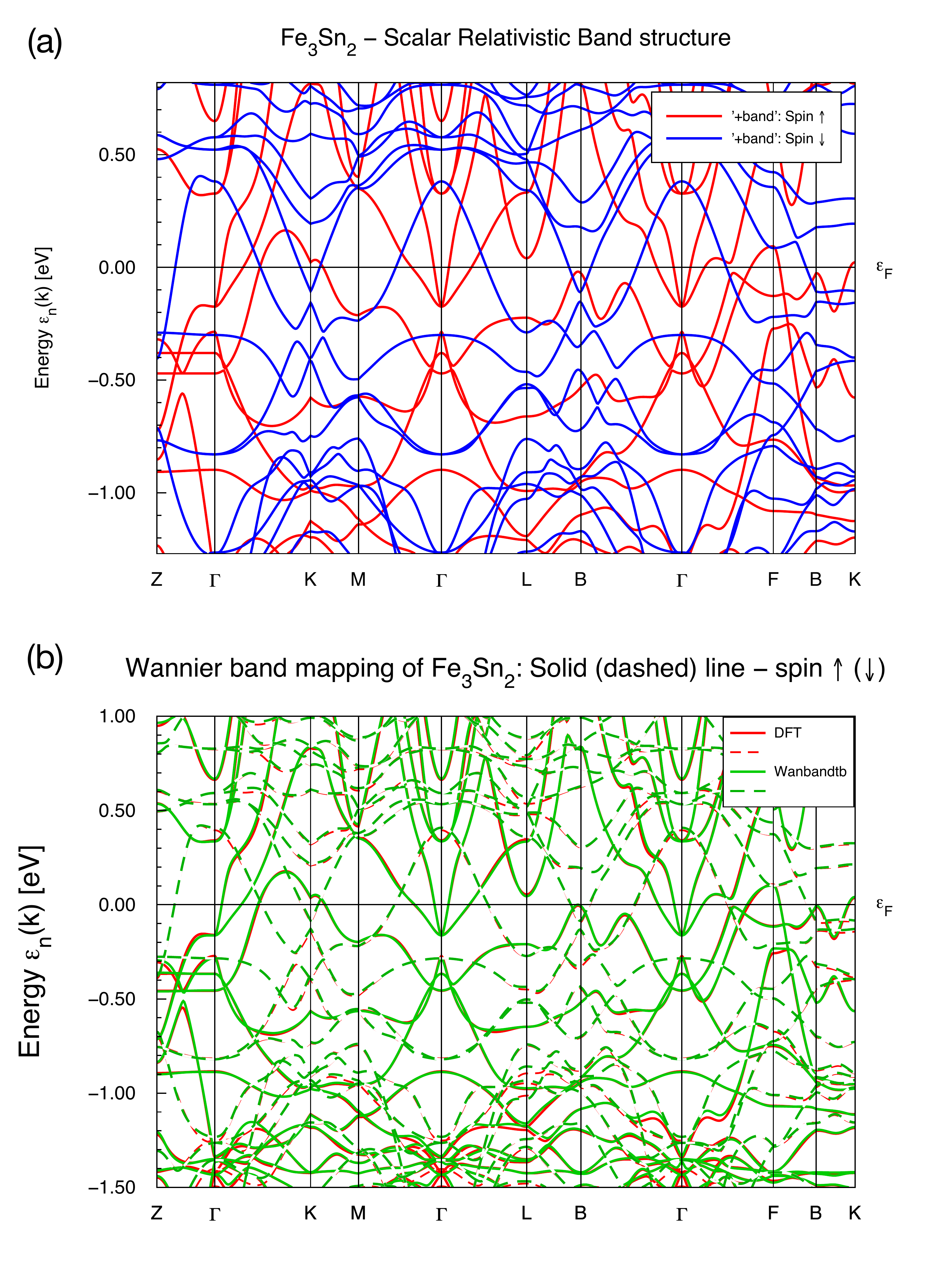}
	\caption{\label{fig-si-FPLO-scalar} Scalar relativistic DFT calculations from FPLO (a) electronic band structure for both spin components (red/blue) (b) Wannier reconstructed bands (green) compared to the full DFT results (red), with two spin components in solid and dashed lines.}
\end{figure}

\begin{figure}[h]
	\includegraphics[width = 0.6\columnwidth]{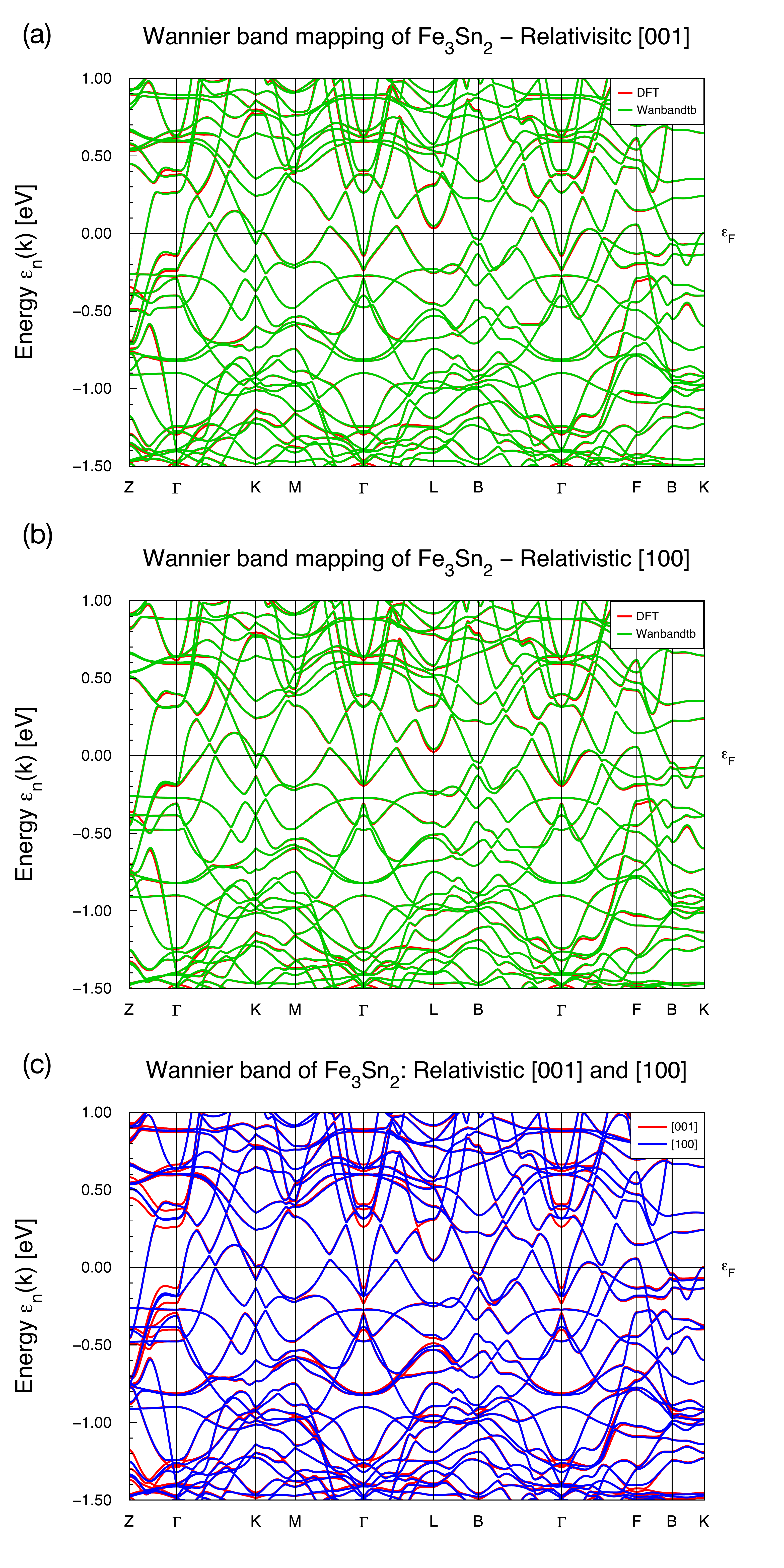}
	\caption{\label{fig-si-FPLO-relativistic} Fully relativistic DFT calculations from FPLO: Wannier reconstructed bands (green) compared to DFT results (red), with the FM order along the (a) [001] (b) [100] directions. (c) Comparison of Wannier electronic band structures with FM order in [001] (red) and [100] (blue) directions.}
\end{figure}

In this section, we compare the electronic structures in the full FPLO DFT calculations and the reconstructed Wannier tight-binding Hamiltonians, as discussed in the method section. The scalar relativistic and fully relativistic results are shown in Fig. \ref{fig-si-FPLO-scalar} and \ref{fig-si-FPLO-relativistic}, respectively.

\FloatBarrier

\subsection{Anomalous Hall conductivity}

\begin{figure}[h]
	\includegraphics[width = 0.7\columnwidth]{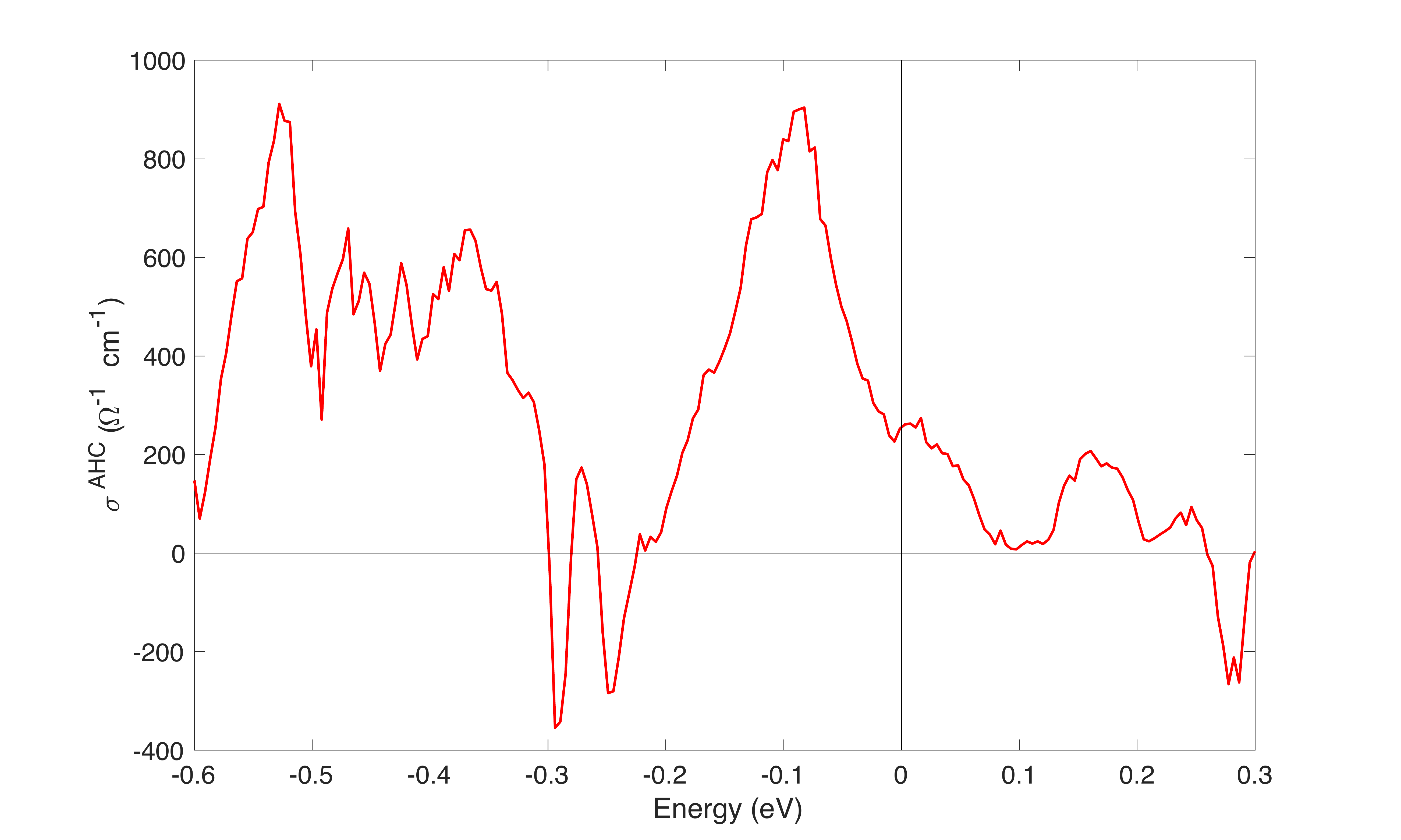}
	\caption{\label{fig-si-sxy} Calculated anomalous Hall conductivity $\sigma_{xy}$ as a function of Fermi level, with FM order along the $c$ axis.}
\end{figure}

\begin{figure}[h]
	\includegraphics[width = 0.95\columnwidth]{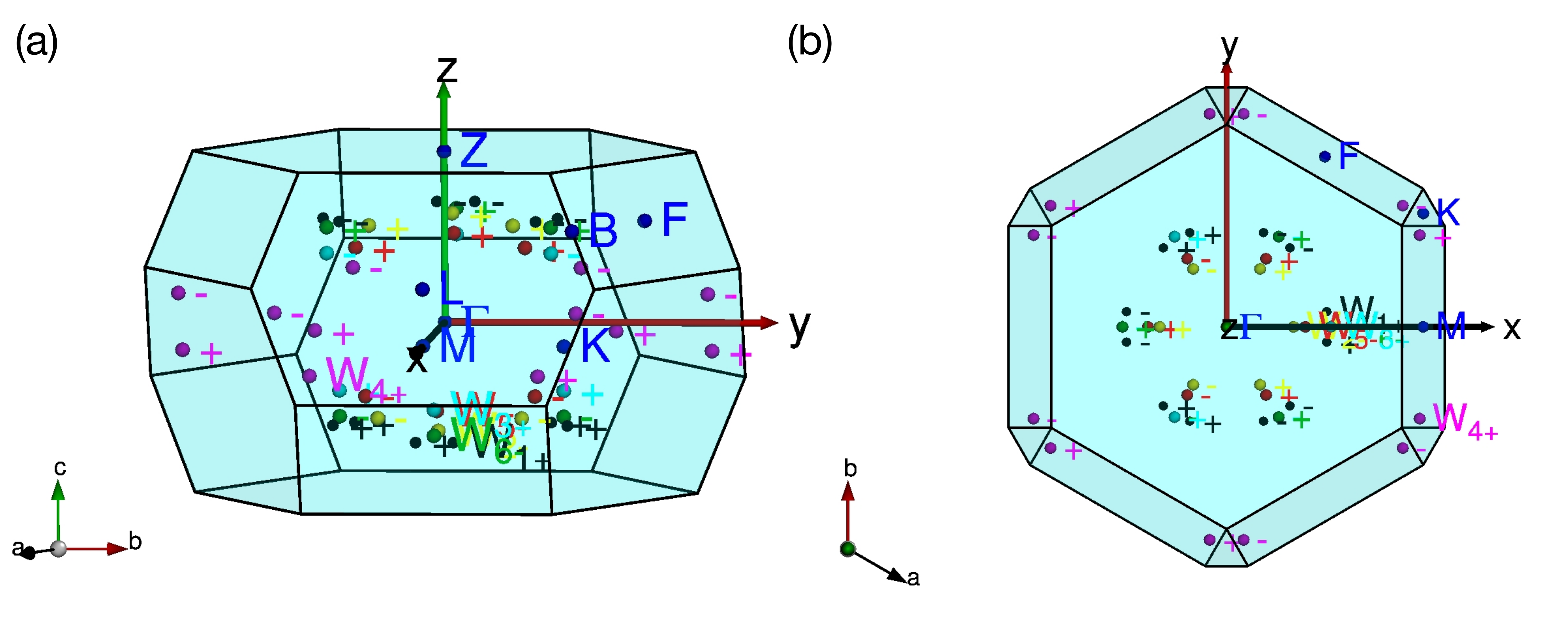}
	\caption{\label{fig-si-weyl} Weyl nodes locations in Fe$_3$Sn$_2$ electronic structure with FM order aligned in [001] direction ($z$ axis). There are six types of Weyl nodes W$_i$ near the Fermi level with the characteristics in Table \ref{tab:weyl_node_001}. The momentum locations of the Weyl nodes are displayed in the (a) side view (b) top view of the 3D BZ. The $\pm 1$ denotes the chiralities of the associated Weyl nodes.}
\end{figure}

With the Wannier tight-binding Hamiltonians, the Berry curvatures of the electronic bands and the integrated anomalous Hall response can be computed~\cite{Berry_numerical_method}. In Fig. \ref{fig-si-sxy}, we show the calculated anomalous Hall conductivity $\sigma_{xy}$ as a function of Fermi level. The features of the total conductivity $\sigma_{xy}$ can be identified with the gap locations of the quasi-two-dimensional massive Dirac fermions with FM order along $c$ axis. For example, the peak around $E = -0.1$ eV corresponds to the massive gap of the upper Dirac structure.

%We can focus on the $\sigma_{xy}$ contributions from the $k$ points close to K/K' points when projected and ignore the contributions from additional band features away from K points. With a cutoff ($xx$ \AA$^{-1}$), the $\sigma_{xy}$ component in the near-K regions is shown in blue curve. As shown in the Berry curvature map in the main text, this partial $\sigma_{xy}$ component shows strong features from the gapped double Dirac cones. For the gap at upper Dirac cone, the $\sigma_{xy}$ is found to be saturated to the quantized value ().

\FloatBarrier

\subsection{Weyl Points Near Fermi Level}

In this section we elaborate on underlying Weyl points in Fe$_3$Sn$_2$ electronic structure. With the Wannier tight-binding Hamiltonians, these Weyl points can be identified using the PYFPLO module of the FPLO package \cite{FPLO}. In Fe$_3$Sn$_2$ crystal with FM order in the $z$ direction, we have identified six types of Weyl points near the Fermi level (see Table \ref{tab:weyl_node_001} for the characteristics).

% within around \pm 50 meV (next one is \pm 90-100 meV)

\begin{table}[h!]
  \centering
 \caption{Characteristics of Weyl points in the Fe$_3$Sn$_2$ electronic structure, when the FM order is along $z$ direction. There are six types of Weyl nodes W$_i$ within energy range 50 meV near the Fermi level. Within each type, the representative $k_i$ positions are also given, in units of \AA$^{-1}$.} 
  \label{tab:weyl_node_001}
  \begin{tabular}{cccccc}
  \hline
    & Energy (meV) & Multiplicity & $k_x$ & $k_y$ & $k_z$ \\
  \hline
  W$_1$ & -33.7 & 12 & -0.183   & 0.028  &  0.160 \\
  W$_2$ & -17.3 & 6 &   -0.123   & 0.0  &  0.148 \\
  W$_3$ & -14.3 & 6 &   -0.191    &    0.0  &  0.111   \\
  W$_4$ & 17.2 & 12 & -0.354  &  0.168 &    0.045  \\
  W$_5$ & 25.1 & 6  &  -0.144  &   0.0  &  0.117 \\
  W$_6$ & 36.1 & 6  &   -0.191    &     0.0  &  0.149  \\
  \hline
  \end{tabular}
\end{table}

%\begin{table}[h!]
%  \centering
% \caption{Characteristics of Weyl points in the Fe$_3$Sn$_2$ electronic structure, when the FM order is along $z$ direction. There are six types of Weyl nodes W$_i$ within energy range 50 meV near the Fermi level. } 
%  \label{tab:weyl_node_0010}
%  \begin{tabular}{ccccccc}
%  \hline
% & W$_1$ & W$_2$ & W$_3$ & W$_4$ & W$_5$ & W$_6$ \\
%  \hline
%Energy (meV) & -33.7 & -17.3 & -14.3 & 17.2 & 25.1 & 36.1 \\
%Multiplicity & 12 & 6 & 6 & 12 & 6 & 6 \\
%  \hline
%  \end{tabular}
%\end{table}

\section{\NoCaseChange{Pressure modification of the helical nodal lines in \ce{Fe3Sn2}}}

\begin{figure}[h]
	\includegraphics[width = 0.8\columnwidth]{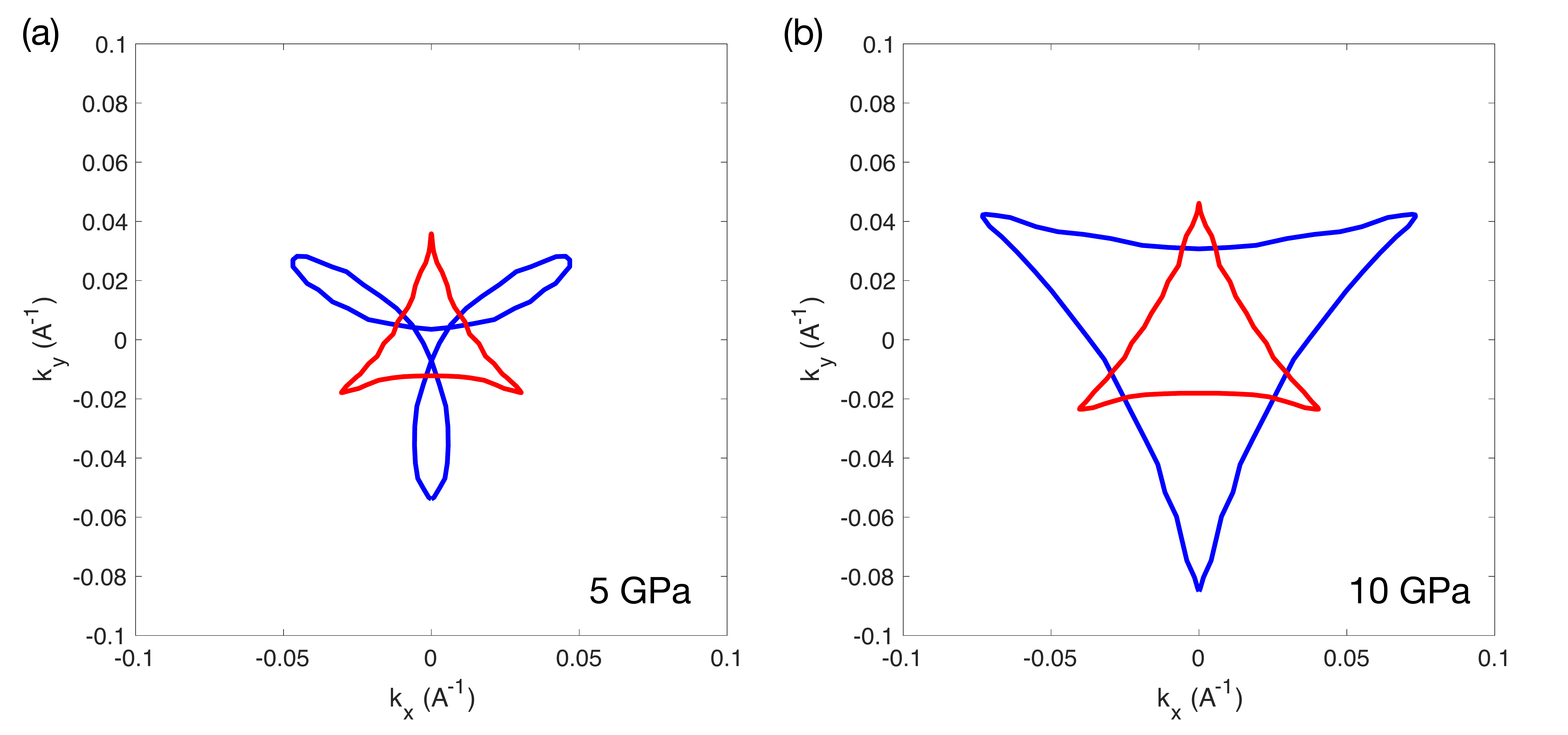}
	\caption{\label{fig-si-pressure} Modification of the helical nodal lines in \ce{Fe3Sn2} by (a) 5 GPa (b) 10 GPa hydrostatic pressure. The blue (red) curves are the helical nodal lines for the lower (upper) Dirac branch. The larger orbit size at higher pressure indicates enhanced inter-layer coupling energy scale $t_{\perp}$ as sheets are closer in distance.}
\end{figure}

In this section, we briefly consider the perturbations on the helical nodal line electronic structure under an external hydrostatic pressure applied to the crystal. This hydrostatic pressure can compress the crystal structure and also reduce the inter-layer distance between kagome sheets. This in turn, modifies the inter-layer coupling energy $t_{\perp}$ between Dirac states on kagome sheets. From the effective models, these coupling energies $t_{\perp}$ are shown to control helical nodal line characteristics, such as the size of the helical nodal line orbit and the Fermi surface topology of the drumhead surface states.

The pressure effects are simulated via density functional theory calculations implemented in the Vienna ab initio simulation package (VASP)~\cite{vasp1,vasp2}. The electronic structure is computed based on the Projector Augmented-Wave~\cite{PAW} pseudopotential formalism and Perdew–Burke-Ernzerhof (PBE)~\cite{pbe} parametrized exchange-correlation energy functional. The \ce{Fe3Sn2} crystals are relaxed under an external hydrostatic pressure of 5 GPa and 10 GPa (up to around 3\% changes in the lattice constants). The helical nodal line structures are shown in Fig. \ref{fig-si-pressure}. The enhanced inter-layer coupling energies $t_{\perp}$ results in the enlarged size of the helix at higher pressure. We note that the shape of nodal line trajectories are DFT code-dependent; nevertheless our results suggest that pressure is an effective tuning parameter of the nodal lines in \ce{Fe3Sn2}.

\section{\NoCaseChange{$\bm{k}\cdot\bm{p}$ expansion for the Dirac structures and the Wannier projection}}

In this section, we discuss the numerical projection procedure to derive the $\bm{k}\cdot\bm{p}$ expansion of the (double) Dirac structure based on the Wannier projection principle~\cite{wannier_review}. Such effective low-energy models can capture the evolution of Dirac structure in the momentum space. We focus on a small region that is confined near $\mathrm{K}$ in the $(k_x,k_y)$ plane while periodically extended along $k_z$. This differs from the conventional $\bm{k}\cdot\bm{p}$ expansion around a single $k$ point, and also the full Wannier transformation which requires the full Brillouin zone sampling~\cite{wannier_review} to derive the localized state.
%To be precise, this is expanded for a small in-plane $(k_x,k_y)$ region centered around K point, at the corner of the hexagonal Brillouin zone. However, it is valid in the extended periodic $k_z$, in the out-of-plane $c$ axis direction. 

The effective modeling here features both approaches. First, we derive a conventional $\bm{k}\cdot\bm{p}$ model with respect to in-plane $(k_x,k_y)$ and identify the proper basis functions to support such expansion. Next, to capture full $k_z$ information, we utilize the Wannier transformation in the out-of-plane direction and derive the localized basis. In short, the final basis functions for the low energy expansion are extended in $x$, $y$ directions, but localized along $z$. In other words, these states are the Dirac states in each layer of kagome sheets, and they form the intricate structure (along $z$) via interplane couplings. In the following, we first motivate the construction of these Dirac states on individual kagome layers (initial projectors) and then derive effective models that account for the energy dispersions and layer hybridizations. We then apply this construction to the ideal AA-BB-CC kagome model and the full \ce{Fe3Sn2} electronic structure.

\subsection{Initial Wannier projector for the Dirac states}

\begin{figure}[h]
	\includegraphics[width = 0.85\columnwidth]{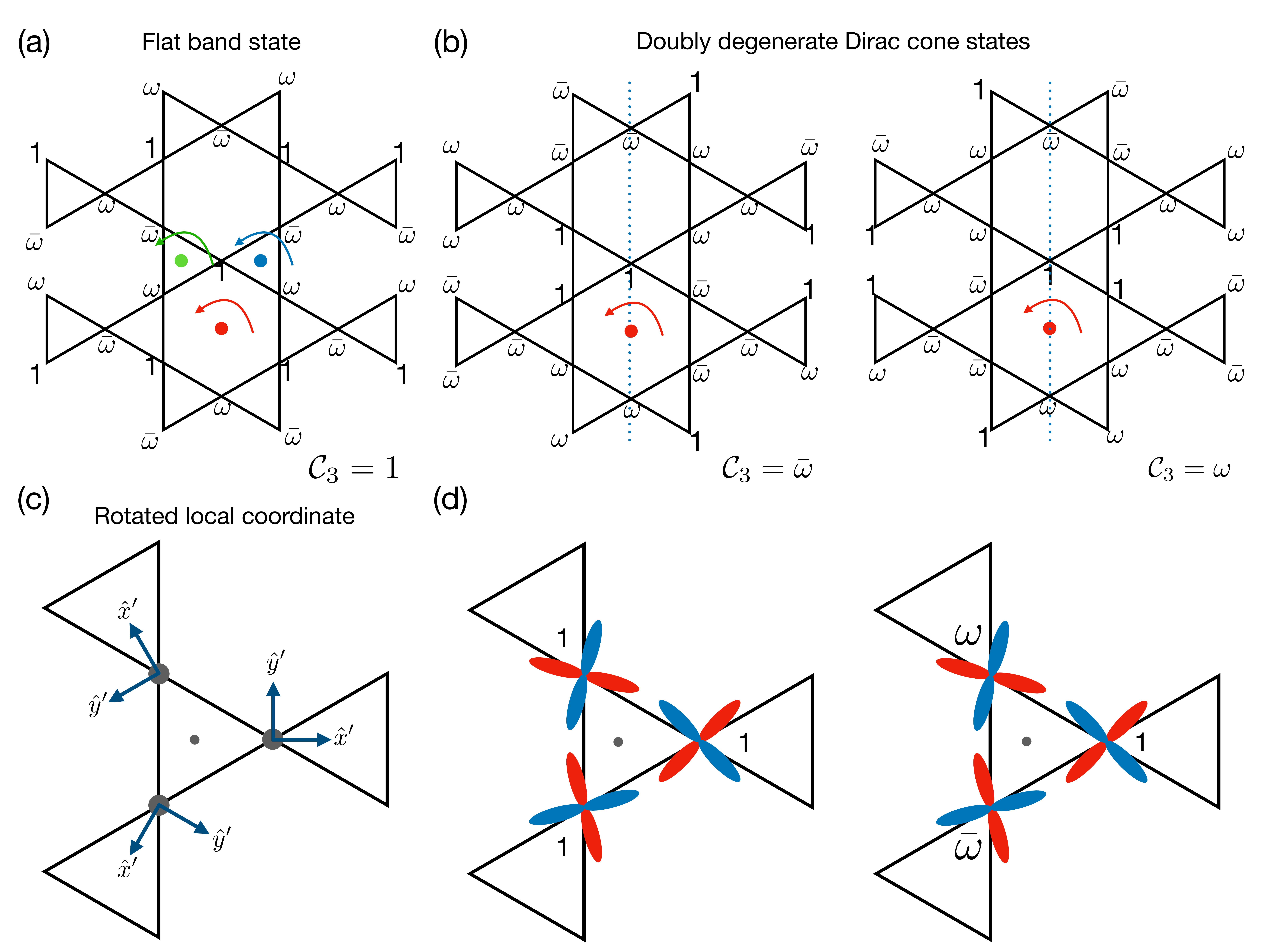}
	\caption{\label{fig-si-kagome-wannier} The symmetry properties of the (a) flat band and (b) doubly degenerate Dirac states in a kagome lattice. The amplitudes for the wavefunction are shown ($\omega=e^{i2\pi/3}$ and $\bar{\omega}=e^{-i2\pi/3}$) on the kagome sites. The red dot at the center of hexagon denotes a three-fold rotation center, and the $\mathcal{C}_3$ symmetry eigenvalues for these states are indicated. Blue and green dots at the centers of triangles are the other choices of three-fold rotation centers. The Dirac doublet states are connected by the mirror symmetry (blue dotted line). (c) The rotated local coordinate system used to define atomic orbital basis (d) The initial Wannier projector used for each kagome sheet with in-plane crystal momentum K. The wave function contains only local $d_{xy}$ orbital component, with the amplitude indicated ($\omega=e^{i2\pi/3}$ and $\bar{\omega}=e^{-i2\pi/3}$).}
\end{figure}

In a simple kagome lattice layer, with sites decorated by the isotropic (single) $s$-wave orbitals, the flat band and the Dirac doublets at K point has the underlying wave function textures as shown in Fig. \ref{fig-si-kagome-wannier}(a,b). From the wave function pattern, the $\mathcal{C}_3$ symmetry eigenvalues at various rotation centers can be inferred. The pairs of the Dirac doublet states are used as the initial basis projection in the double Dirac structure from AA-BB-CC kagome model.

For \ce{Fe3Sn2}, the AA-BB-CC kagome structure is decorated by more complex $d$ orbitals. The multi-orbital nature of the electronic structure can give rise to multiple copies of the Dirac structure in the spectrum, from different set of $d$ orbital orientations. However, for the double Dirac structure in \ce{Fe3Sn2}, the band character is summarized (Table 1 in main text) with local coordinate frame defined in Fig. \ref{fig-si-kagome-wannier}(c). This motivates us to adopt the initial basis functions for projections as shown in Fig. \ref{fig-si-kagome-wannier}(d), compounded with the local $d_{xy}$ orbitals. In the FM (spin-polarized) ordered phase of \ce{Fe3Sn2}, we only consider the minority spin channel that form the double Dirac cones. There are two projected states in a single layer, and two in-equivalent layers, hence a total of four basis states.

In the following, we carry out the procedure and show this projection is non-singular with the initial states. That is, the double Dirac manifold can be captured and expanded by our initial basis functions without obstructions.

\subsection{AA-BB-CC kagome model $\bm{k}\cdot\bm{p}$ expansion and double Dirac structure}

In this section, we discuss the $\bm{k}\cdot\bm{p}$ expansion for the Dirac structures in the AA-BB-CC kagome lattice model (see main text Fig. 4(a)). The full tight-binding model has 6 bands, and the electronic properties for the 4 bands of the Dirac cones are shown in Fig. 4(b-e) under different model parameters. The goal is to derive a 4-band effective model to capture these Dirac states and their nodal lines.

In a given unit cell, there is an AA bilayer unit with two kagome sheets. $L=\pm 1$ denotes the (sub-)layer index. Within each layer, there are two initial states defined at K point for the projections as discussed above, and are denoted as $| \phi^L_i(K,R_z) \rangle$ (with $i=\pm 1$ for the Dirac doublets, and $R_z$ to indicate the bilayer unit index in $z$ stacking direction). These states are extended in $(k_x,k_y)$ and localized in individual kagome sheets along $z$. Now we can perform the Wannier transformation~\cite{wannier_review} along $k_z$ (in-plane momentum fixed at K) for the double Dirac structure. This can be viewed effectively as a transformation in a one-dimensional system with momentum $k_z$. The double Dirac structure in AA-BB-CC with eigenstates $|\Psi_n(K,k_z) \rangle$ defines a projector $\hat{P}(k_z) = \sum_{n} |\Psi_n(K,k_z) \rangle \langle \Psi_n(K,k_z) |$ for the four eigenstates in double Dirac cones. To span the periodic $\hat{P}(k_z)$ manifold, we project by the Fourier-transformed initial states above $|\phi^L_i(K,k_z) \rangle$. We define the new projected states $|\bar{\phi}^L_i(K,k_z) \rangle = \hat{P}(k_z) |\phi^L_i(K,k_z) \rangle$. Although they are not orthonormal, an unitary transformation can be used as in the Wannier transformation~\cite{wannier_review} to obtain an orthonormal basis set $|\Phi^L_i(K,k_z) \rangle$ from the overlap matrix $S=\langle \bar{\phi}^{L'}_{i}(K,k_z) |\bar{\phi}^L_j(K,k_z) \rangle$. An inverse Fourier transform of $|\Phi^L_i(K,k_z) \rangle$ can give the updated Dirac state in the real space $|\Phi^L_i(K,R_z) \rangle$. As with the initial $| \phi^L_i(K,R_z) \rangle$ states above, these new states are extended in the in-plane direction, and exponentially localized in the $z$ direction. We note that the projection is always non-singular in $k_z$, and these four states are able to completely span the double Dirac structure without any obstructions.

After obtaining this proper basis set $|\Phi^L_i(K,k_z) \rangle$, the effective Hamiltonian can be derived by computing the matrix elements $\langle \Phi^L_i(K,R_z) | \hat{H}(K+\delta \vec{k},k_z) | \Phi^{L'}_j(K,R^{'}_{z}) \rangle$. To capture the helical nodal lines, we derive this expansion to include the constant terms and linear terms in $\delta \vec{k}$. These terms capture the on-site energy shifts, inter-layer coupling between single-layer Dirac states, and linear terms for the Dirac dispersion.

Generally, the four band model can be written as 
\begin{equation}
    H(\delta \vec{k},k_z) = \sum_{R_z} (H^0_{R_z}+H^x_{R_z} \delta k_x+H^y_{R_z} \delta k_y)e^{-i R_z k_z}
\end{equation} The $\mathcal{C}_3$ rotation symmetry can be used to derive the selection rules of the matrix elements and simplify the Hamiltonian. One feature of the rhombohedral stacking order is that, the pair of Dirac doubles on a single layer has $\mathcal{C}_3=e^{\pm i 2\pi/3}$ for a rotation axis at the hexagon center. The neighboring layers are relatively shifted which lead to different $\mathcal{C}_3$ symmetry eigenvalue pairs of Dirac doublets. This results in a specific form of non-zero matrix elements. To summarize in a compact form, we use $\bm{\tau}$ to denote the (sub)-layer index and $\bm{\sigma}$ for the Dirac doublets. The effective four band model is shown as in the main text and demonstrates the helical nodal line structure:
\begin{equation}
\begin{split}
\mathcal{H}  = & i\hbar v_F (k_+ \sigma_- -k_- \sigma_+ ) +t_{aa} (e^{i k_z c_1} \tau_+ +e^{-i k_z c_1} \tau_-) \\
& +2 t_{ab} (e^{-i k_z c_2} \tau_+ \sigma_- + e^{i k_z c_2} \tau_- \sigma_+)
\end{split}
\end{equation} We note that the prefactor in $2 t_{ab}$ arises from 6 pairs of inter-layer couplings in an AB bilayer unit.

\subsection{\ce{Fe3Sn2} $\bm{k}\cdot\bm{p}$ expansion and double Dirac structure}

\begin{figure}[h]
	\includegraphics[width = 0.95\columnwidth]{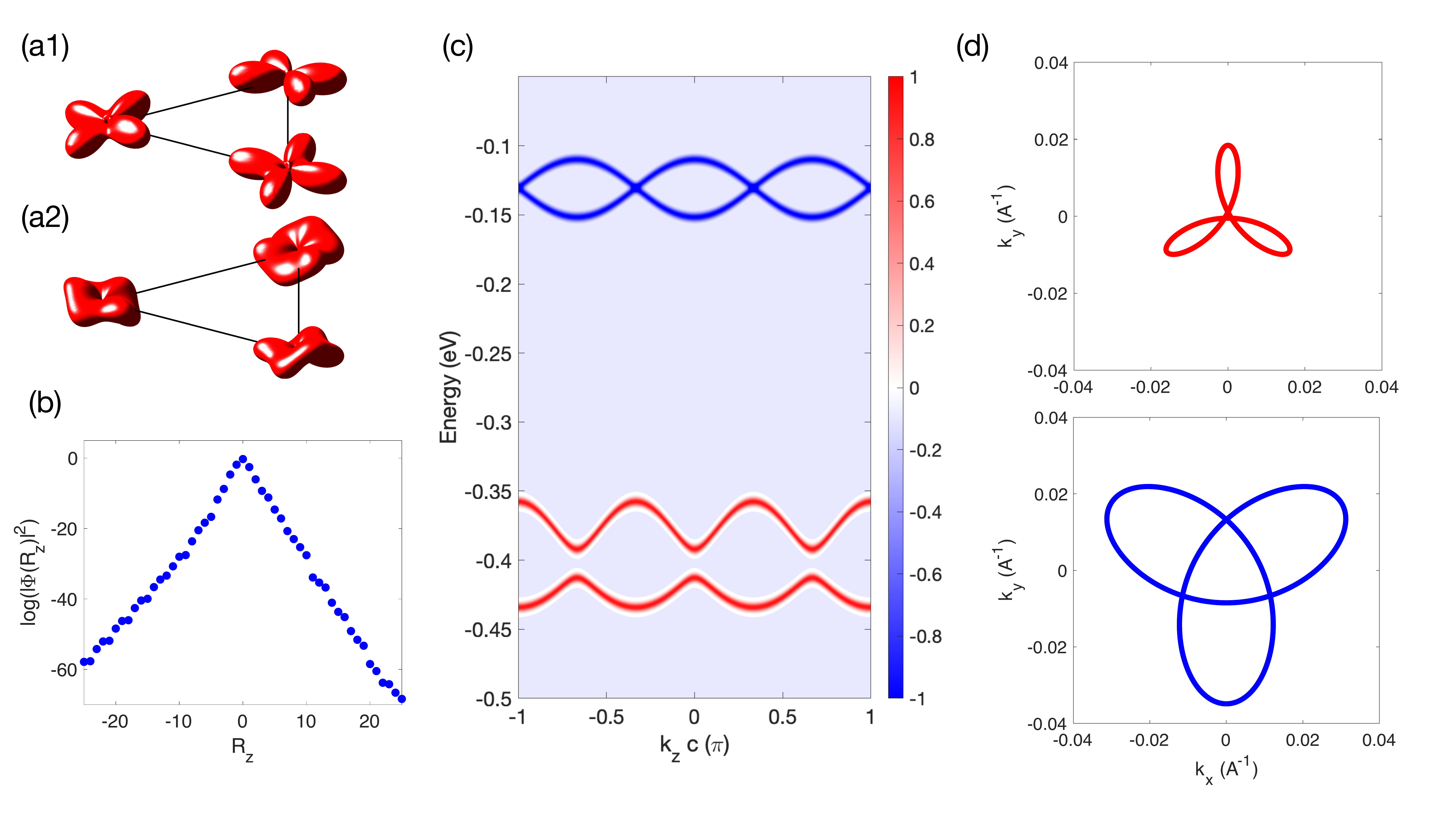}
	\caption{\label{fig-si-fe3sn2-kp} (a1,a2) The two converged Wannier basis states $|\Phi^L_i(K,R_z)|^2$, with $i=\pm 1$, on a kagome sheet for \ce{Fe3Sn2} $\bm{k}\cdot\bm{p}$ expansion of the double Dirac cone structure (b) The localization of these Wannier states, with an exponentially decay in $z$. (c) $\langle \tau_x \rangle$ of the double Dirac states based on the derived $\bm{k}\cdot\bm{p}$ model. This shows the bonding (anti-bonding) combination of the states in the AA-bilayer unit for the lower (upper) Dirac cone, computed along $k_z$ at the projected K point. (d) The top view of the helical upper (red) and lower (blue) nodal lines captured in the $\bm{k}\cdot\bm{p}$ model, which are in good agreement with the full DFT results.}
\end{figure}

For the bulk \ce{Fe3Sn2} crystal, the kagome structure is the same as the AA-BB-CC kagome system above despite the additional Sn spacer layer unit. The double Dirac cone structure also resembles the ideal AA-BB-CC kagome system. Among the five Fe $d$ orbitals, local $d_{xy}$ orbitals are the dominant components in the double cones. With the full Wannier model to describe the \ce{Fe3Sn2} electronic structure, we perform the same projection procedure as above to project the double Dirac cone states. The initial wave function projectors are as the above AA-BB-CC kagome case, and with additional local $d_{xy}$ orbitals attached. Similarly, we find the projection can be carried out as well without any singularity and obstruction. The final projected Wannier basis set for expansion is shown in Fig. \ref{fig-si-fe3sn2-kp}(a), for the Dirac doublets on a single kagome sheet. These Wannier objects also show a strong exponential decay in the wave function weights along $z$ direction away from the center layer as in Fig. \ref{fig-si-fe3sn2-kp}(b). One of the doublets (Fig. \ref{fig-si-fe3sn2-kp} (a1)) preserves better a local $d_{xy}$ component compared to the other (Fig. \ref{fig-si-fe3sn2-kp} (a2)). The stronger out-of-plane component in (a2) may be associated with a $\mathcal{C}_3$ symmetry-allowed hopping channel to its nearest next layer neighbor, which is forbidden by $\mathcal{C}_3$ symmetry for (a1).

%Intuitively, this could be explained by the symmetry allowed couplings between Dirac doublets in a AB-bilayer unit. Only one of the state is allowed to couple to the adjacent kagome layer that is relatively shifted, due to the transformation of $\mathcal{C}_3$ symmetry representation. Because of this coupling, the orbital content is more mixed compared to the other state with symmetry-forbidden couplings.

With this proper basis set, the effective Hamiltonian can be derived up to the linear term in $\delta \vec{k}$. The results are elaborated in Methods. One feature we find from the effective model projection is that, the coupling in the AA bilayer unit $t_{aa}$ is larger than the coupling in the AB bilayer unit $t_{ab}$ between Dirac states. As the result, the double Dirac cone splitting arises from the bonding / anti-bonding of the Dirac cone states in an AA bilayer unit (despite the Sn spacer layer in between) as in Fig. \ref{fig-si-fe3sn2-kp}(c) for $\langle \tau_x \rangle$. Within this low-energy Hamiltonian, the helical nodal lines can be captured as shown in Fig. \ref{fig-si-fe3sn2-kp}(d).

% tAB vs tAA
We expand on the observation $t_{ab} < t_{aa}$, despite a smaller interlayer distance in an AB bilayer unit, compared to an AA bilayer unit. This appears at first sight contradictory to larger microscopic atomic orbital couplings derived from the (full) Wannier Hamiltonian based on the DFT calculations, as we show in Table \ref{tab:micro_wan_tab_taa}. Hereafter we elaborate on the origin of the reduced effective $t_{ab}$ coupling in the effective $\bm{k}\cdot\bm{p}$ Hamiltonian. The main difference between the (full) Wannier model and the effective $\bm{k}\cdot\bm{p}$ model is that, the irrelevant states are integrated out and the renormalized basis states are derived within the effective model. This process naturally mixes states of different orbital types, locations, and renormalizes the coupling matrix elements. In the derived Wannier state, the electron density is spread out across several nearby kagome layers. To be precise, in an AB bilayer, the wavefunction for the Wannier state $\Phi^{l}$ centered at $l=$ A, B can be decomposed into layer-resolved components $\psi^{l}_k$ at layer $k=$ A, B. Generally, $\Phi^{l}$ can spread further exponentially as shown in Fig. \ref{fig-si-fe3sn2-kp}(b), but we focus only on the components within an AB bilayer unit. In particular, the Wannier state centered on the kagome A layer in an AB bilayer has a non-zero density weight about 1/5 on the neighboring kagome B layer compared to the density weight on the kagome A layer. In the context of orbital textures, the dominant components for weights $\psi^{A}_A$ on kagome A layer are the in-plane $d$ orbitals ($d_{xy}$ and $d_{x^2-y^2}$), as used in the initial projector. In contrast, the weights $\psi^{A}_B$ on the kagome B layer are mostly out-of-plane $d_{yz}$ orbitals. These out-of-plane $d$ orbital contents can be viewed as induced from enhanced atomic couplings involving out-of-plane orbital sectors compared to the fully in-plane ones in an AB bilayer, as tabulated in Table \ref{tab:micro_wan_tab_taa}.

Thanks to the layer-dependent orbital texture of the effective basis states, the matrix element for the effective $t_{ab}$ contains contributions from different channels between these components within the AB bilayer $\sum_{i,j} \langle \psi^{B}_i | \hat{H}(K) | \psi^{A}_j \rangle$.  For the dominant contributions and channels within an AB bilayers, the numerical numbers are shown in Table \ref{tab:eff_wan_tab}, and a cancellation is observed between these $\langle \psi^{B}_i | \hat{H}(K) | \psi^{A}_j \rangle$ channels which reduces the effective coupling $t_{ab}$ by one order of magnitude compared to the microscopic values. We note that numerically, there are additional cancelling terms for wavefunction components spread into further kagome and Sn spacer layers, which further suppresses $t_{ab}$.

On the other hand, the microscopic interplane coupling across A-A stacking in Table \ref{tab:micro_wan_tab_taa} appear small compared that across A-B stacking. However, we noticed the effective coupling can be mediated via the intermediate Sn $s/p$ orbitals from the spacer layer unit, and the states can efficiently tunnel and hybridize. This leads to the enhanced effective $t_{aa}$ couplings, and the bonding / anti-bonding states across the Sn layer for the double cone structure.

In summary, the effective basis states as the composite objects strongly renormalize the coupling parameters from the microscopic values. The geometric stacking pattern and the spacer layer electronic properties can affect the effective theory. It would be interesting to consider the design of these factors and the implications on the electronic structure.

%The effective $t_{ab}$ coupling between such composite objects has the contributions from different channels in between these Wannier components. A cancellation is observed between these channels which reduces the effective coupling.

%\begin{table}[h!]
%  \centering
% \caption{The microscopic $d$-orbital couplings for the shortest interlayer Fe pair in an AB bilayer unit. The orbitals are grouped into three sectors, $d_{xy}$/$d_{x^2-y^2}$, $d_{xz}$/$d_{yz}$ and $d_{z^2}$. We tabulate the matrix norm $||h||_F=\sqrt{\sum_{i,j} |h_{ij}|^2 }$ for the coupling matrix, in units of meV.} 
 % \label{tab:micro_wan_tab}
 % \begin{tabular}{c|ccc}
 % \hline
  %$|| h_{AB} ||_F$ & $d_{xy}$/$d_{x^2-y^2}$ & $d_{xz}$/$d_{yz}$ & $d_{z^2}$ \\
 % \hline
 % $d_{xy}$/$d_{x^2-y^2}$ & 115 & 217 & 243 \\
 % $d_{xz}$/$d_{yz}$ & 217 & 478 & 370 \\
 % $d_{z^2}$ & 243 & 370 & 50 \\
 % \hline
 % \end{tabular}
%\end{table}

%\begin{table}[h!]
 % \centering
% \caption{The microscopic $d$-orbital couplings for the shortest interlayer Fe pair (approximately on top of each other) in an AA bilayer unit across Sn spacer layer. The orbitals are grouped into three sectors, $d_{xy}$/$d_{x^2-y^2}$, $d_{xz}$/$d_{yz}$ and $d_{z^2}$. We tabulate the matrix norm $||h||_F=\sqrt{\sum_{i,j} |h_{ij}|^2 }$ for the coupling matrix, in units of meV.} 
 % \label{tab:micro_wan_taa}
 % \begin{tabular}{c|ccc}
 % \hline
 % $|| h_{AA} ||_F$ & $d_{xy}$/$d_{x^2-y^2}$ & $d_{xz}$/$d_{yz}$ & $d_{z^2}$ \\
 % \hline
 % $d_{xy}$/$d_{x^2-y^2}$ & 30 & 7 & 4 \\
 % $d_{xz}$/$d_{yz}$ & 7 & 45 & 1 \\
 % $d_{z^2}$ & 4 & 1 & 82 \\
 % \hline
 % \end{tabular}
%\end{table}

\begin{table}[h!]
  \centering
 \caption{The microscopic $d$-orbital couplings for the shortest interlayer Fe pairs in the AB bilayer unit, and the AA bilayer unit across the Sn spacer layer (the pair of atoms are approximately on top of each other). The orbitals are grouped into three sectors, $d_{xy}$/$d_{x^2-y^2}$, $d_{xz}$/$d_{yz}$ and $d_{z^2}$. We tabulate the Frobenius matrix norm $||h||_F=\sqrt{\sum_{i,j} |h_{ij}|^2 }$ for the coupling matrix between these sectors, in units of meV.} 
  \label{tab:micro_wan_tab_taa}
  \begin{tabular}{cc}
  \begin{tabular}{|c|ccc|}
  \hline
  $|| h_{ab} ||_F$ & $d_{xy}$/$d_{x^2-y^2}$ & $d_{xz}$/$d_{yz}$ & $d_{z^2}$ \\
  \hline
  $d_{xy}$/$d_{x^2-y^2}$ & 115 & 217 & 243 \\
  $d_{xz}$/$d_{yz}$ & 217 & 478 & 370 \\
  $d_{z^2}$ & 243 & 370 & 50 \\
  \hline
  \end{tabular}
  \begin{tabular}{|c|ccc|}
  \hline
  $|| h_{aa} ||_F$ & $d_{xy}$/$d_{x^2-y^2}$ & $d_{xz}$/$d_{yz}$ & $d_{z^2}$ \\
  \hline
  $d_{xy}$/$d_{x^2-y^2}$ & 30 & 7 & 4 \\
  $d_{xz}$/$d_{yz}$ & 7 & 45 & 1 \\
  $d_{z^2}$ & 4 & 1 & 82 \\
  \hline
  \end{tabular}
  \end{tabular}
\end{table}

\begin{table}[h!]
  \centering
 \caption{Here we decompose the contributions of \ce{Fe3Sn2} effective $t_{ab}$ in $\bm{k}\cdot\bm{p}$ model, between Wannier functions $| \Phi^{l} \rangle$ in an AB bilayer unit. The layer-resolved components within the bilayer are $| \psi^{l}_k \rangle$ for Wannier center $l$ decomposed at layer $k$. A cancellation is observed between these $\langle \psi^{B}_i | \hat{H}(K) | \psi^{A}_j \rangle$ channels which reduces the effective coupling $t_{ab}$. The couplings are in units of meV.} 
  \label{tab:eff_wan_tab}
  \begin{tabular}{c|cc}
  \hline
  $\langle \psi^{B}_i | \hat{H}(K) | \psi^{A}_j \rangle$  & $| \psi^{A}_{A} \rangle$ & $| \psi^{A}_{B} \rangle$ \\
  \hline
  $\langle \psi^{B}_{A}|$ &  -130  & 131  \\
  $\langle \psi^{B}_{B}|$ &  170 &  -130  \\
  \hline
  \end{tabular}
\end{table}

\end{document}